\documentclass[12pt]{article}
\usepackage{oldgerm}
\usepackage{yfonts}
\usepackage{euler}
\usepackage{graphics}
\usepackage{bm}
\usepackage{graphicx}
\usepackage{amsmath}
\usepackage{amssymb}
\usepackage{amstext}
\usepackage{amscd}
\usepackage{amsfonts}
\usepackage{appendix}
\usepackage{wasysym}
\usepackage{epstopdf}
\usepackage{color}
\usepackage{hyperref}
\DeclareMathAlphabet{\EuFrak}{U}{euf}{m}{n}
\DeclareMathAlphabet{\EuScript}{U}{eus}{m}{n}

\newcommand{\nd}{\noindent}

\newcommand{\be}{\begin{equation}}
\newcommand{\ee}{\end{equation}}
\newcommand{\ben}{\begin{eqnarray}}
\newcommand{\een}{\end{eqnarray}}

\title{{\bf Boltzmann and Tsallis statistical approaches to study Quantum corrections at large distances and clustering of galaxies }}
\author{{M. Hameeda$^{1,2,a}$,Q. Gani$^{1,b}$, B. Pourhassan$^{3,4,c}$,}\\
{\texttt{ \rm{M.C.Rocca$^{5,6,7,d,e}$}}}\\
\small{$^1$ Department of Physics, Cluster University Srinagar, 190001 India}\\
\small{$^2$ Inter University Centre for Astronomy and Astrophysics , Pune India}\\
\small{$^3$ School of Physics, Damghan University,}\\
\small{ P. O. Box 3671641167, Damghan, Iran}\\
\small{$^4$ Canadian Quantum Research Center 204-3002 32 Ave Vernon, }\\
\small{BC V1T 2L7 Canada}\\
\small{$^5$ Departamento de F\'{\i}sica,
Universidad Nacional de La Plata,}\\
\small{$^6$ Departamento de Matem\'{a}tica,
Universidad Nacional de La Plata,}\\
\small{$^7$ Consejo Nacional de Investigaciones Cient\'{\i}ficas
y Tecnol\'{o}gicas}\\
\small{(IFLP-CCT-CONICET)-C. C. 727, 1900 La Plata -
Argentina}\\
\small{\texttt{\rm{$^{a}$hme123eda@gmail.com, $^{b}$qudsiagani6@gmail.com, $^{c}$b.pourhassan@du.ac.ir,}}}\\
\small{\texttt{\rm{$^{d}$rocca@fisica.unlp.edu.ar,$^{e}$mariocarlosrocca@gmail.com, }}}}

\date{\today}

\begin{document}

\maketitle

\begin{abstract}
\nd  Gravity is so different from other fundamental forces that it is now essentially treated as a non-fundamental force of entropic origin. A number of good studies have been carried out in this direction. Quantum gravity has also significantly improved our understanding by combining gravity well with quantum physics. However, there are still many impediments to our understanding especially in the limits of extreme. The effects of quantum gravity start appearing on the scene at Planck length which is the smallest length in nature idealized so far \cite{ali, mead, amati}. While as this study incorporates a model which is valid for potential energy corrections at small distances but we have also given a bold try to use it confidently for the corrections at very large distances as well.  The model uses two techniques namely Boltzmann and Tsallis statistical approaches to explore the thermodynamics within the ambit of brane world model giving its further modified version. We have computed partition function by using both Boltzmann and Tsallis statistical approaches and then used it to study thermodynamics of the brane world model. We have analyzed both analytically and graphically, the thermodynamic quantities like Helmholtz free energy and specific heat. The thermodynamic stability of the model is also discussed depending on the number of galaxies.\\\\
\nd
\nd

\end{abstract}

\newpage

\tableofcontents

\newpage

\renewcommand{\theequation}{\arabic{section}.\arabic{equation}}

\section{Introduction}

\nd
To understand the properties of a physical system, we need to perturb it to observe the changes. According to Heisenberg's uncertainty principle, the change in any two canonically conjugate quantities cannot be measured simultaneously with precision. Quantum gravity, on the other hand, suggests that if one observable cannot be measured with certainty so cannot be the other. However, the probability distribution of unperturbed systems  like the gravitating systems which we are interested in, are measured in terms of Generalized Uncertainty Principle (GUP).  For a physical system
\begin{equation}
\label{eq1.1}
 [q_{j}, p_{k}]  = q_{j}(-\hbar i)(\frac{\partial}{\partial q_{k}}) - (-\hbar i)(\frac{\partial}{\partial q_{k}})(q_{j})
\end{equation}
which reduces to
\begin{equation}
\label{eq1.2}
\sigma q_{j}. \sigma p_{k} \ge \frac{1}{2} \hbar \delta_{jk}.
\end{equation}
If j$\neq$k then $q_{j}$ and $p_{k}$ can be measured simultaneously to any degree of accuracy.  Now a fundamental relation between the force of gravity and the minimal length in nature was found in certain studies \cite{mead} String theory suggests that gravitational force in reality could be a strong force but it appears weak since it has to escape through a large  number of extra dimensions. The size of the dimension is essentially the point of debate because no interactions  can take place at distances smaller than the size of the string \cite{amati}. Our aim is to use the effect of GUP on gravity as an entropic force and find the modifications of potential in the  Newtons law. 

\nd
A detailed derivation which includes modification of the Newtonian potential as a series expansion has been gone through in the RS model with a detailed discussion of leading order term for short distances where also, a bold claim has been made about the gravity being $5$-dimensional at short distances \cite{pet}.
Different views on theories with extra dimensions have been put forward in different studies. It is pertinent to mention that the Kaluza-Klein picture considers the radius of the extra dimension to be the Planck length and millimeter-sized extra dimensions suggested by Arkani-Hamed et al. \cite{ark, ark1} and an infinite extra dimension proposed by Randall and Sundrum \cite{ran}.
The braneworld scenario thus hints at the possibility of observing higher dimensional effects on gravitational force at sub-millimeter scales. Studying modifications to the conventional cosmology and astrophysics from the perspective of braneworld models and subsequent comparison of the predictions to the observational data especially that of cosmic microwave background can be the indirect ways of testing the existence of extra dimensions\cite{thesis}. An investigation of the effects of density perturbations in a brane world universe with dark radiation has also been carried out\cite{bur}.
In one of the relevant works, it has been asserted that the observable universe behaves like a $3$-dimensional membrane surface in a higher-dimensional spacetime, which can be the source of new perspectives in the study of cosmology \cite{roy}.
 That the brane world cosmology can bring new twists in structure formation, clustering of galaxies, inflation and the ideas for dark energy, gives a sufficient motivation for pursuing the present study. In present work, we analyze the effects of brane world on the partition function by taking the short distance modification term in the gravitational potential  and evaluate the corresponding thermodynamics using the Boltzmann and Tsallis statistical approaches.

\nd 
Black holes are an important platform where the study of gravity can be connected to thermodynamics. These behave as black bodies while emitting photons and the temperature of peak emission can be found using Wien's law.  In this context, Planck scale effect is already studied \cite{ali}. We want  to go far ahead to see the effects at much larger distances say Hubbles length or so. This is reasonable to assume because our observational Universe does have a huge radius. This will be a totally new theory to explain these corrections. The idea is that deformation of quantum mechanics  at short distance  (from quantum gravity ) produces these $1/R^{2} $ corrections. Now we assume that quantum like effects occur at very large distances too. So, we have three phases to nature\\
\nd 1] Short distance where quantum effects occur due to $\hbar$ corrections. \\
2] Intermediate distance where classical physics can be applied (when both $\hbar$ and $\alpha$ effects can be neglected) and\\
3] Very large distance where quantum effects reoccur due to $\alpha$ corrections.\\

\nd
The large scale structure of our universe is largely due to clustering of galaxies. The clustering of galaxies, their distribution functions and the resulting thermodynamic properties  have been thouroughly analyzed using Newtonian gravity both for point masses \cite{sas}; \cite{sas2}; \cite{sas84} and for extended masses. \cite{fmh},\cite{ahm06}, \cite{ahm10}; \cite{Hameeda2} 
The modification in the gravity action can lead to the variations in  the gravitational potential in the low-energy limit and the modified potential reduces to the Newtonian one on the Solar scale as well. It has been analyzed in \cite{mil1}; \cite{mil2}; \cite{mil3} that the modified gravitational potential could fit galaxy rotation curves even in absence of dark matter or dark energy. This in fact has, open a possibility to tie an analogy between the moderations due to the modified Newtonian potential and the dark energy models.

\nd The clustering of galaxies has been pursued by considering Newtonian modified potential of a brane world model by using usual techniques of statistical mechanics have been discussed \cite{Hameeda}. The corrections to the Newtonian potential are because of the super-light modes in the brane world models. It may be noted that the Newtonian potential also gets altered from different approaches. The essential to mention here include non-commutative geometry \cite{nic}; \cite{gre}, minimal length in quantum gravity \cite{ali}, f(R) gravity \cite{noj}, \cite{cap}, $\Lambda$CDM \cite{Ham20} and the entropic force
\cite{maj}. Work has been done to see the impact of modified potentials on thermodynamic properties of gravitational clustering \cite{4}, \cite{Hameeda2}; \cite{Hameeda}. 
\nd The effect of the cosmological constant on the clustering of galaxies and the influence on distribution of galaxies has been studied \cite{1}; \cite{1b}. Essential to mention here is the phenomenological Tohline–Kuhn modified gravity approach to the problem of dark matter \cite{kuh1},\cite{toh1}; \cite{toh2}.
Some of the aspects of the nonlocal modifications of Newtonian gravity and linearized gravitational waves have been studied in detail \cite{non}. The problems with action principles for nonlocal theories in regard to the  symmetric issues of kernel has been discussed in Hehl and Mashhoon (2009b) and a classical nonlocal generalization of Einstein's theory of gravitation has been  developed \cite{bah,bah1,bah2,bah3}.
The field equation of nonlocal gravity leads to the possibility that nonlocal gravity may replicate dark matter \cite{mas}. To investigate whether nonlocal gravity is fulfilling the capability of solving the problem of large-scale cosmological structure formation is the important task with the cosmologists. 
Gravitational clustering has been analyzed in terms of partition function which may diverge because of the treatment of galaxies as point particles in the extended structure of the universe \cite{sas84}. However, this divergence has been removed by introducing a softening parameter which also moderates the thermodynamic fluctuations \cite{ahm02, ahm06, Hameeda2}. 
A recent work has been dedicated to study the clustering of galaxies by pursuing the nonlocal aspect of gravity as the nonlocal extension of general relativity \cite{nonbt}.  
Exploring the possibility of relevance of GR at small scales where quantum effects are important has suffered due to lack of experimental support, on the other hand studying quantum effects at large scales is becoming an interesting area of investigation because of the increasing hopes of future  experimental supports \cite{rich}. The possibility for the existence of maximum length was raised in context of the cosmological particle horizon as the maximum measurable length in the Universe \cite{epjc}.
The extension of the standard quantum field theory proposes that effects related to the multivalued nature of the physical fields could be seen at large spatial scales \cite{kir}. The logarithmic behavior of the gravitational potential at large distances could appear as a result of nontrivial properties of the vacuum state in Modified Field Theory \cite{kir1}.
In the subsequent sections, the thermodynamics of the system of galaxies has been studied by using the partition function with the large distance brane corrections. We are considering $N$ number of galaxies with pairwise gravitational interaction between them  and performing thermodynamic calculations using Tsallis and Boltzman statistical approaches.
The results are important and novel for being the large distance quantum modifications to the gravity. This is a kind of first of its studies of gravitational clustering with quantum effects which reoccur at very large distances due to $\alpha$ corrections. This study can be considered a proposal that at large distances quantum effects reoccur and can influence the thermodynamic properties and the distribution of galaxies in the expanding uiverse!!!  

\setcounter{equation}{0}

\section{Boltzmann's divergence free partition function of an interacting system}

\begin{figure}[h!]
 \begin{center}$
 \begin{array}{cccc}
\includegraphics[width=70 mm]{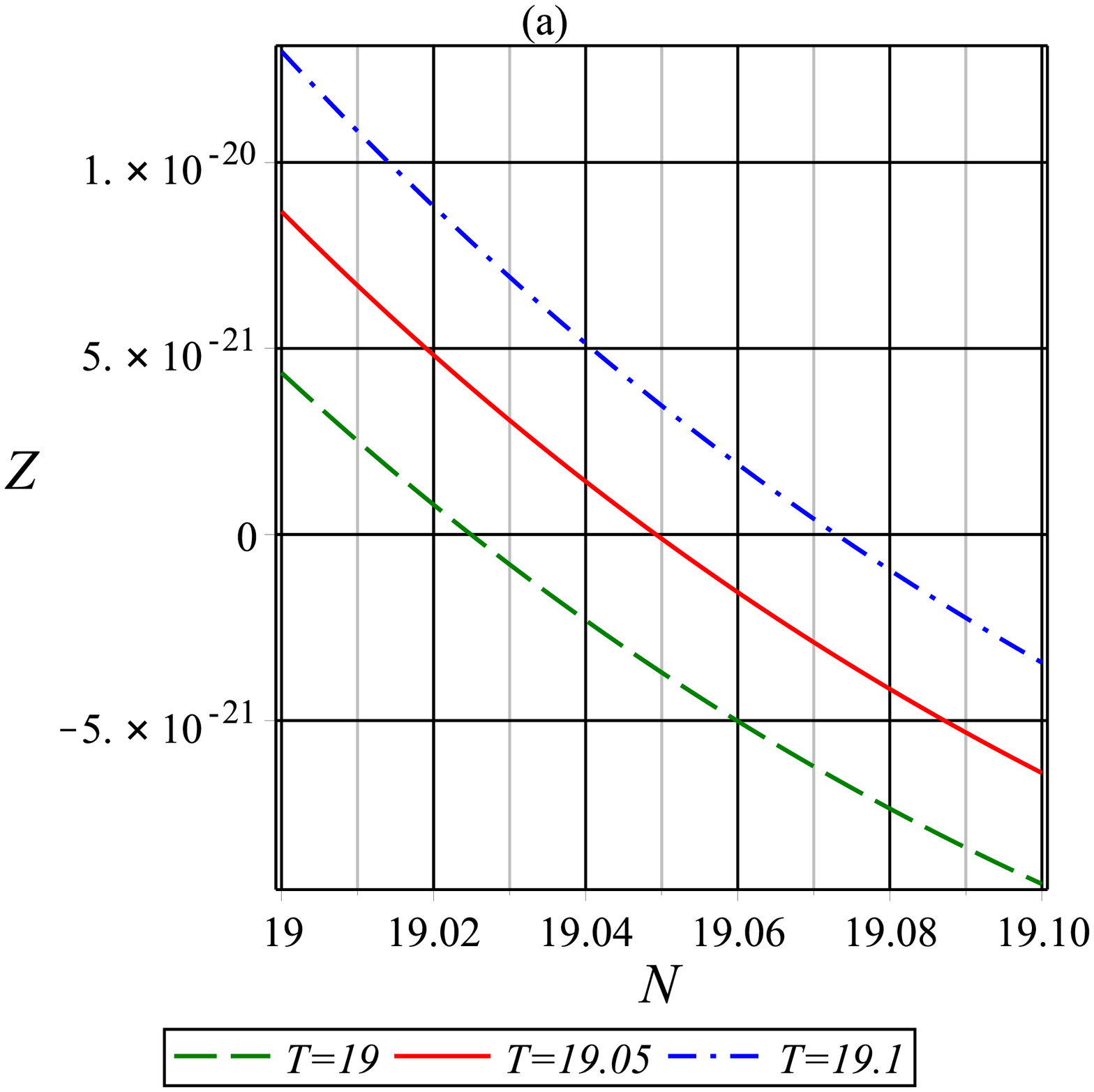}\includegraphics[width=70 mm]{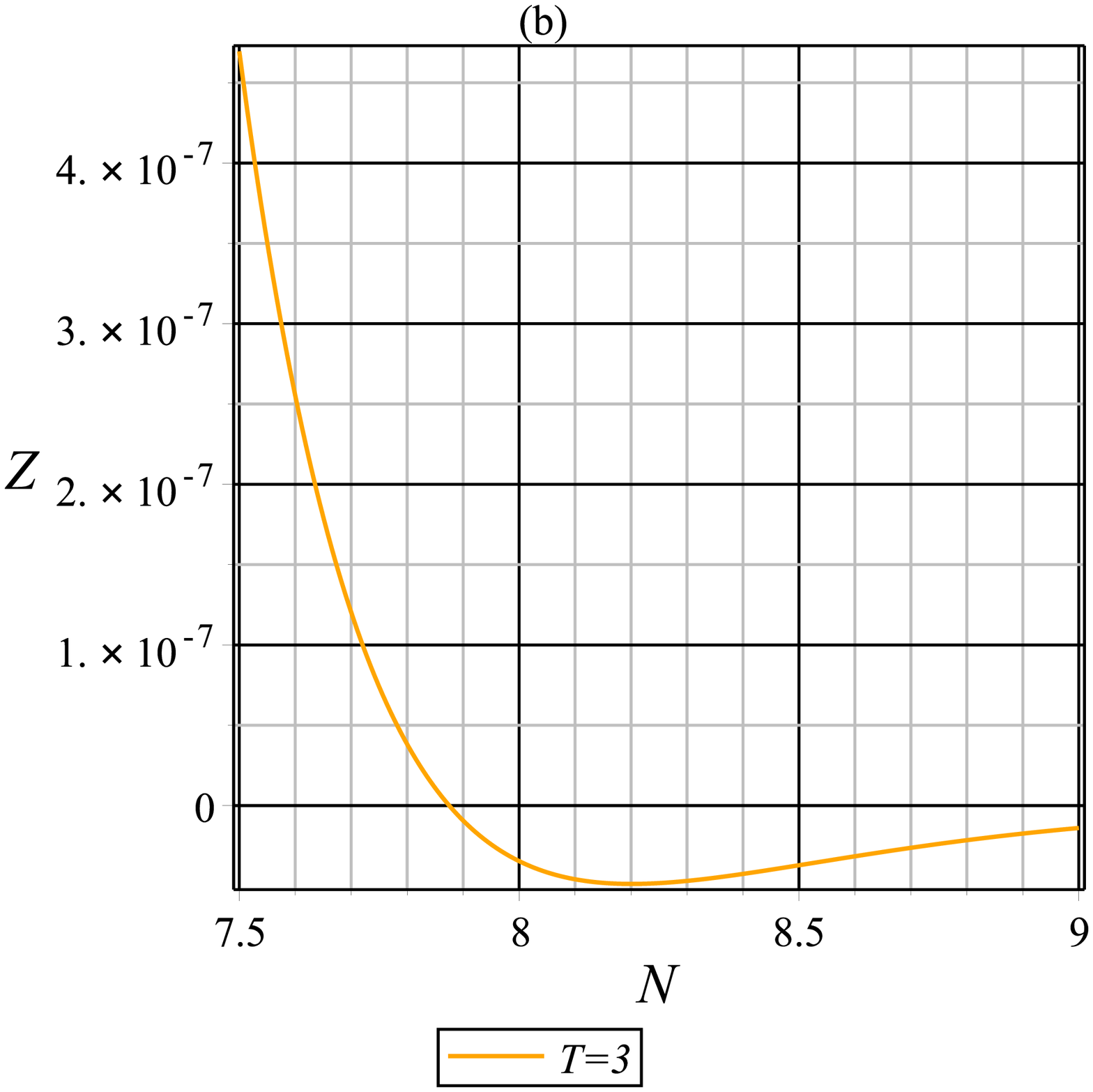}
 \end{array}$
 \end{center}
\caption{Partition function in terms of $N$ for the unit values of the parameters.}
 \label{fig0}
\end{figure}

There is an established connection between gravitation and thermodynamics and all physical systems can be described in terms of their degrees of freedom. Consequently we can evaluate partition functions. The general partition function of a system of $N$ particles of mass $m$ interacting through the modified gravitational potential which together with potential energy  $\Phi$,  can be written as
\begin{eqnarray}
\label{eq2.1}
{\cal Z}_\nu&=& \frac{1}{N!}\int d^{\nu}pd^{\nu}r
\exp\biggl(-\frac{\sum_{i=1}^{N}\frac{p_{i}^2}{2m}+\Phi(r_{1}, r_{2}, r_{3},
\dots, r_{N})}{T}\biggr),
\end{eqnarray}
\begin{equation}
\label{eq2.2}
\phi_{i,j}=-\frac{N(N-1)Gm^2}{ 2r_{ij}}\left(1+\frac{k}{ r_{ij}}\right),
\end{equation}
Consequently, the partition function ${\cal Z}$ in $\nu$ dimensions can be written as
\begin{eqnarray}
\label{eq2.3}
{\cal Z}_\nu&=&\frac{1}{N!}\int\limits_{-\infty}^{\infty}d^\nu x\int\limits_{-\infty}^{\infty}d^\nu p
\exp\left[\beta\left(\frac {N(N-1)Gm^2} {2r}\left(1+\frac{k}{ r}\right)-\frac {Np^2} {2m}\right)\right]\nonumber\\
&=&\frac{1}{N!}\left(\frac {2\pi^{\frac {\nu} {2}}} {\Gamma\left(\frac {\nu} {2}\right)}\right)^2
\int\limits_0^{\infty}r^{\nu-1}dr\exp\left(\beta\frac {N(N-1)Gm^2} {2r}\left(1+\frac{k}{r}\right)\right)\nonumber\\
 &\times&\int\limits_0^{\infty}p^{\nu-1} dp\exp\left(-\frac {Np^2} {2m}\right)
\end{eqnarray}
Now, we may use the following standard integral \cite{gr}
\begin{equation}
\label{eq2.5}
\int\limits_0^\infty dx x^{\nu-1}e^{-\beta x^2-\gamma x}=
(2\beta)^{-\frac {\nu} {2}}\Gamma(\nu)e^{\frac {\gamma^2} {8\beta}}
D_{-\nu}\left(\frac {\gamma} {\sqrt{2\beta}}\right)
\end{equation}
where $D_\nu$ is the parabolic cylinder function and we can rewrite  it as
\begin{equation}
\label{eq2.6}
\int\limits_0^\infty dx x^{\nu-1}e^{\beta x^2+\gamma x}=
(-2\beta)^{-\frac {\nu} {2}}\Gamma(\nu)e^{-\frac {\gamma^2} {8\beta}}
D_{-\nu}\left(-\frac {\gamma} {\sqrt{-2\beta}}\right)
\end{equation}
This modifies the two integrands of equation (\ref{eq2.3}) as
\[\int\limits_0^{\infty}r^{\nu-1}dr\exp\left(\beta\left(\frac {N(N-1)Gm^2} {2r}\right)\left(1+\frac{k}{r}\right)\right)=\]
\begin{equation}
\label{eq2.7}
\frac{k}{\beta N(N-1)Gm^2}\phi\left(\frac {1+\nu} {2},\frac {3} {2},-\frac {\beta N(N-1)Gm^2k^2} {8}\right)
\end{equation}
and
\begin{equation}
\label{eq2.8}
\int\limits_0^{\infty}p^{\nu-1}dr\exp\left(-\beta\left(\frac {Np^2} {2m}\right)\right)=
\frac {\Gamma\left(\frac {\nu} {2}\right)(2m)^{\frac {\nu} {2}}} {2(\beta N)^{\frac {\nu} {2}}}
\end{equation}
Both the above integrals do not diverge and hence can be expressed at $\nu=3$.
Thus
\begin{equation}
\label{eq2.9}
{\cal Z}=\frac {\sqrt{2}\pi^{\frac {5} {2}}} {N!}\left(\frac {m} {\beta N}\right)^{\frac {3} {2}}
\frac {k} {\beta N(N-1)Gm^2}
\phi\left(2,\frac {3} {2},-\frac {\beta N(N-1)Gm^2k^2} {8}\right),
\end{equation}
where $\phi(a,b,c)$ is confluent hypergeometric function. In order to have physical partition function, it should be positive. It is clear that $N=1$ leads to divergence and we should have $N\geq2$ as required for the interacting system. However, we find that there is an upper limit for $N$ to have physical partition function. We call it $N_{max}$ which is dependent on temperature. In that case $Z\geq0$ if $N\leq N_{max}$ which is illustrated by Fig. \ref{fig0}. We can define a critical $N$ as $T=N_{c}=19.05$ (see Fig. \ref{fig0} (a)), so that $T<N_{max}$ if $N<N_{c}$ and $T>N_{max}$ if $N>N_{c}$. Hence, in the case of $N=N_{c}$ we have $T=N_{c}=N_{max}$. The conclusion is that the large $N$ limit of physical system needs high temperature whileas the number of galaxies  is limited at low temperatures. For example at $T=3$ we can find the physical system with $1<N<7$, so $N_{max}\approx7$.\\

\subsection{Thermodynamic properties using Boltzmann statistics}

\begin{figure}[h!]
 \begin{center}$
 \begin{array}{cccc}
\includegraphics[width=70 mm]{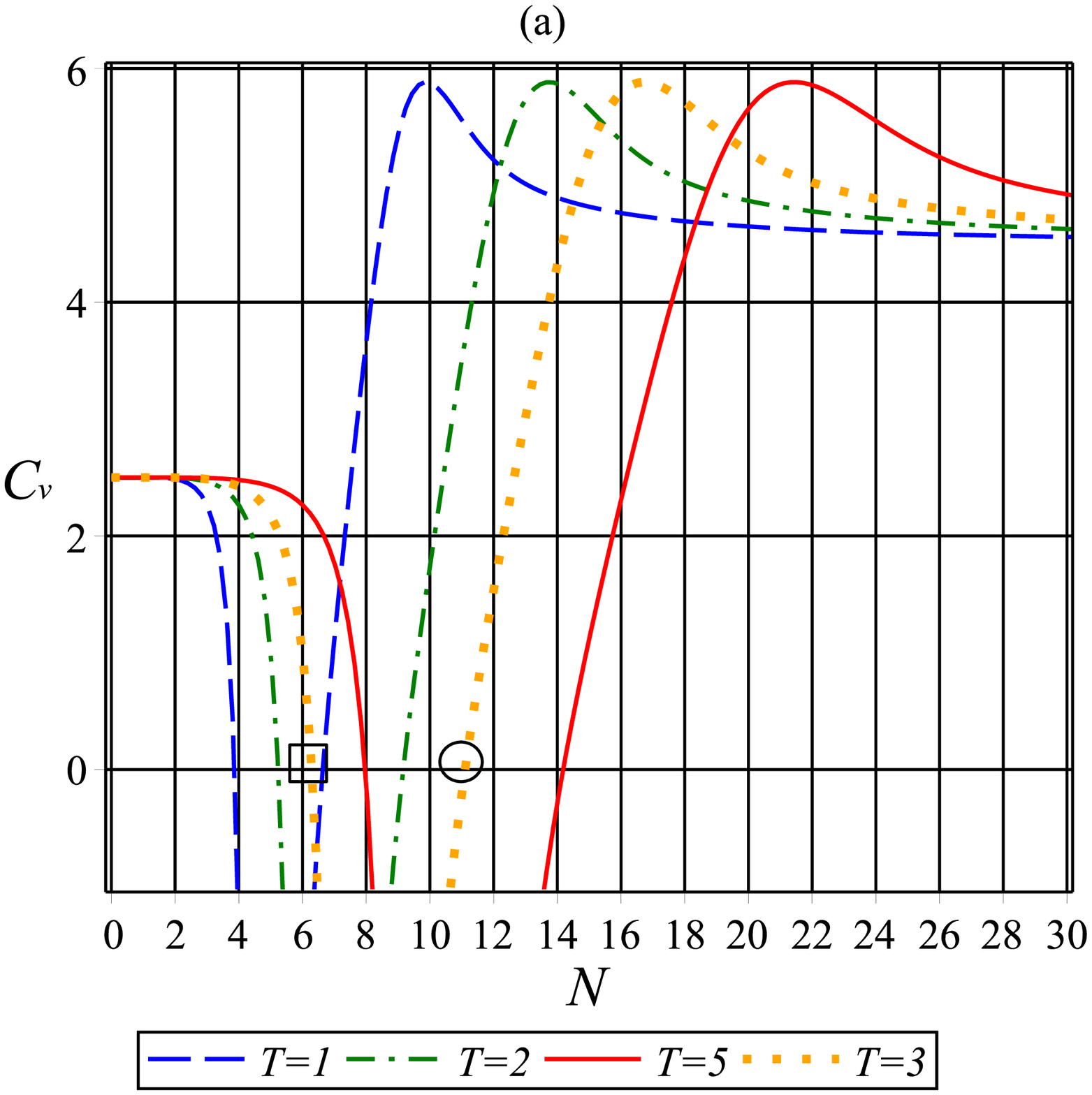}\includegraphics[width=70 mm]{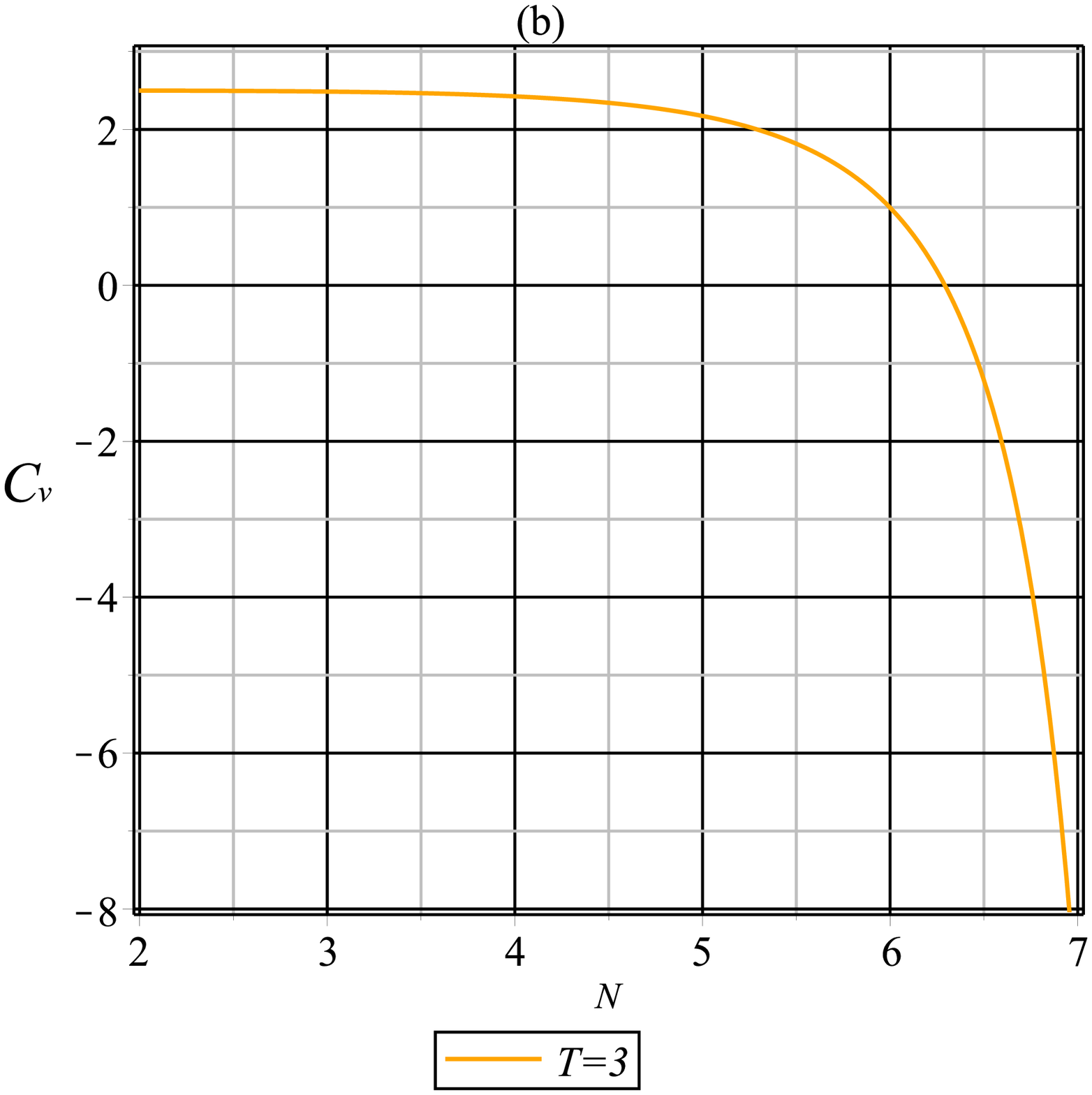}
 \end{array}$
 \end{center}
\caption{Specific heat in terms of $N$ for the unit values of the parameters.}
 \label{fig1}
\end{figure}

In this section, we have obtained the expressions corresponding to different thermodynamically important parameters such as internal energy, Helmholtz free energy, specific heat, entropy, chemical potential and pressure in terms of the modified gravitational potential.
\newline The internal energy $\cal U $  can be expressed  in terms of partition function $\cal Z$ as
\begin{equation}\label{eq3.1}
<{\cal U}>=-\frac {1} {{\cal Z}}\frac {\partial{\cal Z}} {\partial\beta}
\end{equation}
where $\beta$ has its usual meaning,  $\beta = \frac{1}{k_{B}T}$.\\
Now, using the value of $\cal Z$  from the previous section, we get
\begin{equation}
\label{eq3.2}
<{\cal U}>=\frac {5} {2\beta}\left[1+\frac {8\alpha_0}{15}\frac{\phi\left(3,\frac {5} {2},-\alpha_0\right)}{\phi\left(2,\frac {3} {2},-\alpha_0\right)}\right]
\end{equation}
where
\begin{equation}
\label{eq3.3}
\alpha_0=\frac {\beta N(N-1)Gm^2k^2} {8}
\end{equation}
 Further, the specific heat is  obtained from
\begin{equation}
\label{eq3.4}
C_v=\left(\frac{\partial U}{\partial T}\right)_V
\end{equation}
which with the substitution of $\cal U$ from equation (\ref{eq3.2}) comes out to be
\begin{equation}
\label{eq3.5}
C_v=\frac{5k_B}{2}\left[1-\frac{32}{45}{\alpha_0}^2\left(\frac{\phi\left(3,\frac {5} {2},-\alpha_0\right)}{\phi\left(2,\frac {3} {2},-\alpha_0\right)}\right)^2+\frac{48}{75}{\alpha_0}^2\frac{\phi\left(4,\frac {7} {2},-\alpha_0\right)}{\phi\left(2,\frac {3} {2},-\alpha_0\right)}\right].
\end{equation}
In Fig. \ref{fig1} we have drawn specific heat in terms of $N$ for some values of the low temperatures. There are two cases of large and small $N$ limits where the specific heat yields to a constant. Also, there are two critical $N$ where specific heat is zero, which we label as $N_{-}$ (represented by square in dotted orange line of Fig. \ref{fig1}) and $N_{+}$ (represented by circle in dotted orange line of Fig. \ref{fig1}). Consequently we are obtaining an analytical expression for $N_{-}$ which is following ahead. It is found that the value of critical $N$ increases by increasing temperature. For the $N\geq N_{+}$ the specific heat is positive and we see the appearance of Schottky anomaly where the heat capacity rises to a maximum \cite{1907.00248, 1905.00539}, and which is then reduced to a positive constant. Hence for the $N\geq N_{+}$, system is in the stable phase. Such Schottky anomaly may be due to some radiative process or particle creation of a physical system. However, as is discussed already, we  do not have  a physical system in this range  because of  a negative partition function.\\
On the other hand for the case of $N_{-}<N<N_{+}$ the specific heat is negative and the system is in an unstable phase. There is however other region corresponding to $N\leq N_{-}$ where the specific heat is positive which is indeed the domain of our physical system. Comparing Fig. \ref{fig0} with Fig. \ref{fig1} (a) it is clear that $N_{max}>N_{-}$. Thus there exist both regions  in the given system, stable and unstable, which are illustrated better in Fig. \ref{fig1} (b). The stability of  physical system  is subject to the condition  $N\leq N_{-}$, where $N_{-}$ is smallest real positive root of $C_v=0$.\\

\begin{figure}[h!]
 \begin{center}$
 \begin{array}{cccc}
\includegraphics[width=65 mm]{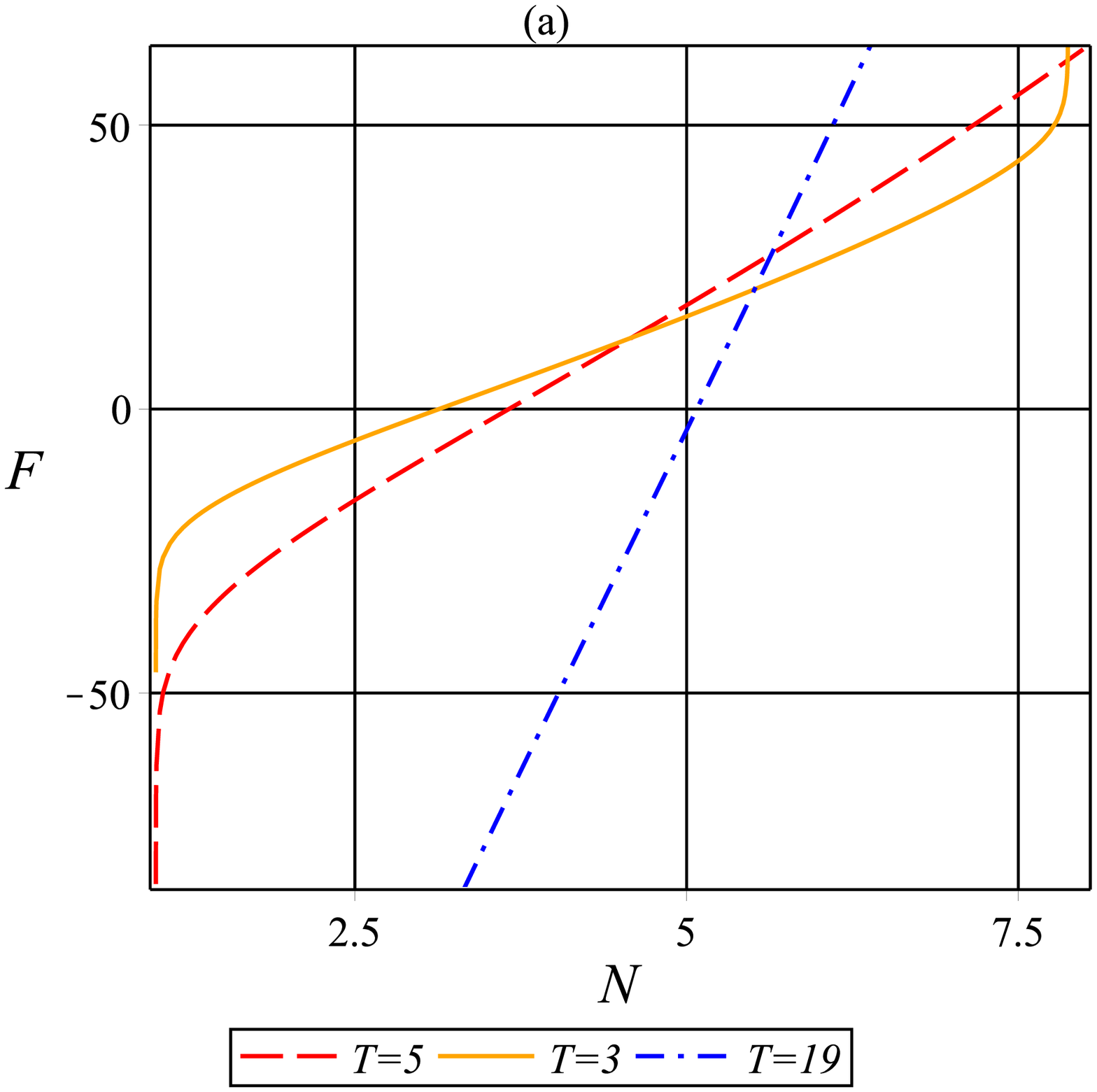}\includegraphics[width=65 mm]{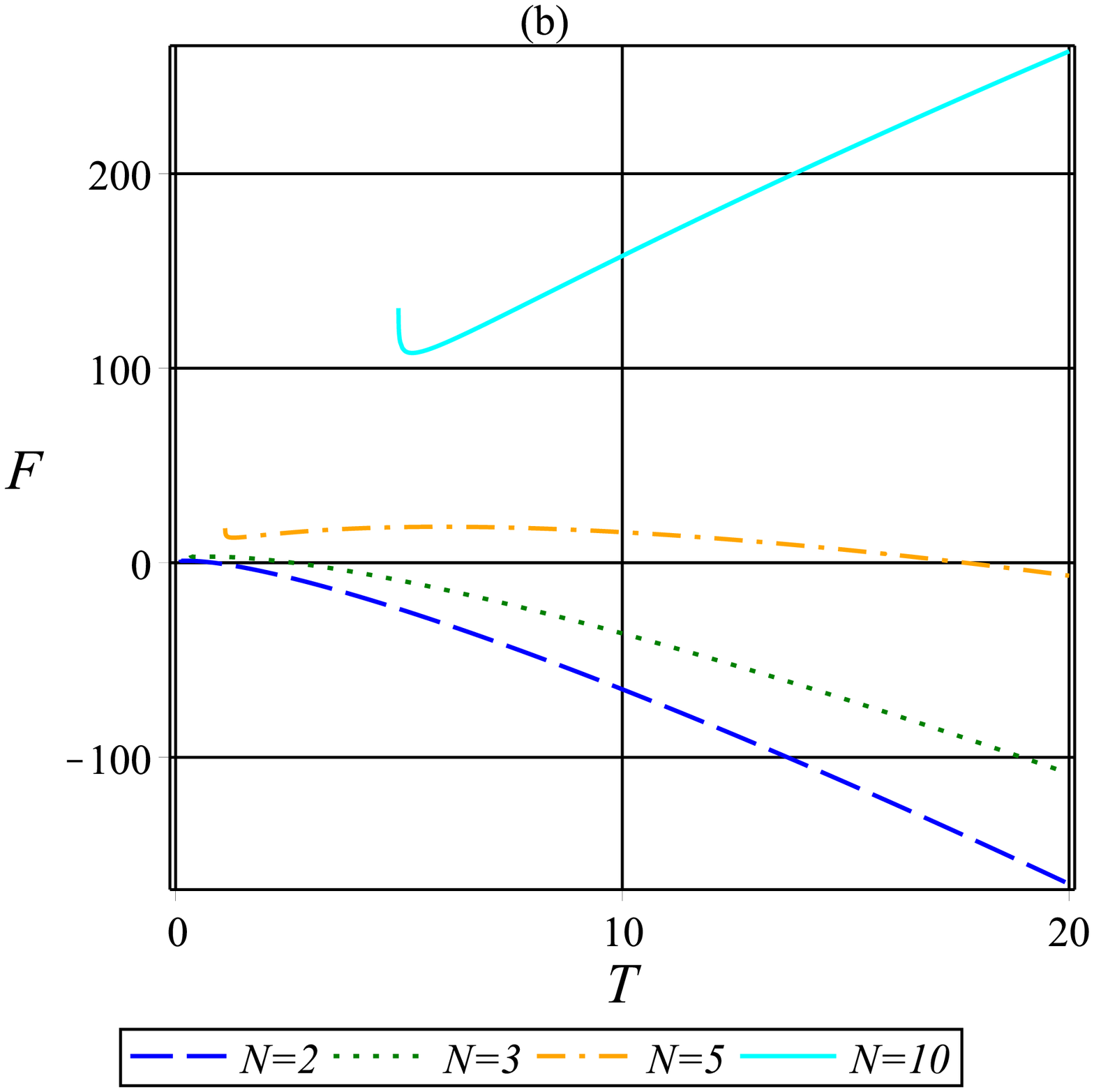}
 \end{array}$
 \end{center}
\caption{Helmhotz free energy in terms of (a) $N$, and (b) $T$ for the unit values of the parameters.}
 \label{fig2}
\end{figure}

Assuming $\alpha_{0}$ as small parameter, we can expand hypergeometric function to obtain roots of equation, $C_v=0$ . Neglecting $\mathcal{O}(\alpha_{0}^{4})$ give us the following
\begin{equation}\label{eq3.55}
\frac{2}{5}-\frac{8}{45}\alpha_{0}^{2}-\frac{32}{189}\alpha_{0}^{3}=0.
\end{equation}
Inserting $\alpha_{0}$ from the equation (\ref{eq3.3}) and including higher order terms, we get the following approximate solution,
\begin{equation}\label{eq3.555}
N_{-}\approx\frac{1}{2}(1+\sqrt{1+44T})
\end{equation}
Similarly, the Helmhotz free energy $F$ is obtained using
\begin{equation}
\label{eq3.6}
F=-\frac{1}{\beta}\ln {\cal Z}.
\end{equation}
Substituting for $\cal Z$, we get
\begin{equation}
\label{eq3.7}
F=-\frac{1}{\beta}\ln\left(\frac {\sqrt{2}\pi^{\frac {5} {2}}} {N!}\left(\frac {m} {\beta N}\right)^{\frac {3} {2}}
\frac {k} {\beta N(N-1)Gm^2}
\phi\left(2,\frac {3} {2},-\alpha_0\right)\right).
\end{equation}
We have drawn Helmhotz free energy in terms of both $N$ and $T$ in Fig. \ref{fig2} to see typical behavior of free energy. It shows two divergent points. The larger one corresponds to the divergence of specific heat which is illustrated by Fig. \ref{fig1}. We can also see that small $N$ yields to negative free energy while larger $N$ system has positive free energy which corresponds to the unphysical case already discussed. In the case of larger $N$, we can see a minimum for the Helmhotz free energy (see solid cyan line of Fig. \ref{fig2} (b)). The dash dotted blue line of Fig. \ref{fig2} (a) corresponds to the case of $T=N_{c}$ with the similar general behavior with other temperatures.

\begin{figure}[h!]
 \begin{center}$
 \begin{array}{cccc}
\includegraphics[width=80 mm]{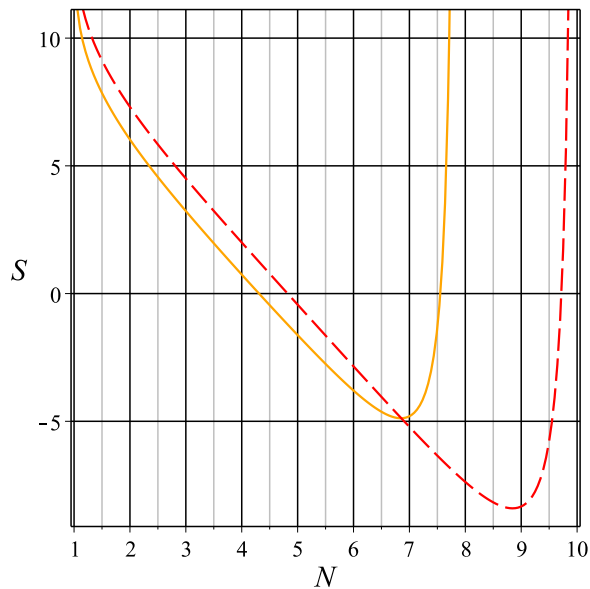}
 \end{array}$
 \end{center}
\caption{Entropy in terms of $N$ for the unit values of the parameters.}
 \label{figS}
\end{figure}

The entropy $S$ is conventionally obtained from
\begin{equation}
\label{eq3.8}
TS=U-F
\end{equation}
which with the substitution for $\cal U$ and $F$ gives
\[S=\frac {5k_B} {2}\left[1+\frac {8\alpha_0}{15}\frac{\phi\left(3,\frac {5} {2},-\alpha_0\right)}{\phi\left(2,\frac {3} {2},-\alpha_0\right)}\right]\]
\begin{equation}
\label{eq3.9}
+k_B\ln\biggl[\frac {\sqrt{2}\pi^{\frac {5} {2}}} {N!}\left(\frac {m} {\beta N}\right)^{\frac {3} {2}}
\frac {k} {\beta N(N-1)Gm^2}
\phi\left(2,\frac {3} {2},-\alpha_0\right)\biggr]
\end{equation}
A further analysis of the entropy shows the constraints on the number of galaxies to have physical situation with positive partition function and positive entropy. We find two points where $S=0$. These are labeled as $N_1$ (represented by square in solid orange line of Fig. \ref{figS}) and $N_2$ (represented by circle in solid orange line of Fig. \ref{figS}). Comparing it with Fig. \ref{fig1}, we find that $N_{1}<N_{-}$ and $N_{-}<N_{2}<N_{+}$. Hence, the physical region with positive entropy decreases as $N\leq N_{1}$.

\begin{figure}[h!]
 \begin{center}$
 \begin{array}{cccc}
\includegraphics[width=80 mm]{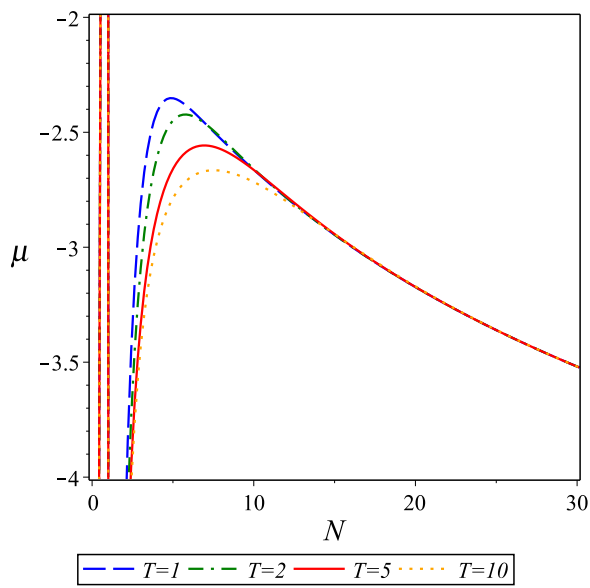}
 \end{array}$
 \end{center}
\caption{Chemical potential in terms of $N$ for the unit values of the parameters.}
 \label{fig3}
\end{figure}

The next important parameter is that of the chemical potential, expressed as
\begin{equation}
\label{eq3.10}
\mu=\left(\frac{\partial F}{\partial N}\right)_T,
\end{equation}
which modifies to
\begin{equation}
\label{eq3.11}
\mu=-\left(\frac{7N-5}{2N(N-1)}\right)-\ln N-\frac{4(2N-1)}{3N(N-1)}\phi\left(2,\frac {3} {2},-\alpha_0\right).
\end{equation}
We have drawn chemical potential in terms of $N$ in Fig. \ref{fig3} to see typical behavior. We can see the divergence  at small $N$, while there is a peak which corresponds to the divergence of specific heat. We can see that chemical potential yields to negative constant at large $N$ limit which is independent of temperature.

 In a similar way, the following relation is used to calculate the Pressure-Volume product,
\begin{equation}
\label{eq3.12}
PV=\frac{2N}{3}U,
\end{equation}
which in terms of $\phi$ comes out to be
\begin{equation}
\label{eq3.13}
PV=\frac {5N} {3\beta}\left[1+\frac {8\alpha_0}{15}\frac{\phi\left(3,\frac {5} {2},-\alpha_0\right)}{\phi\left(2,\frac {3} {2},-\alpha_0\right)}\right]
\end{equation}

\subsection{Probability distribution function}

\begin{figure}[h!]
 \begin{center}$
 \begin{array}{cccc}
\includegraphics[width=80 mm]{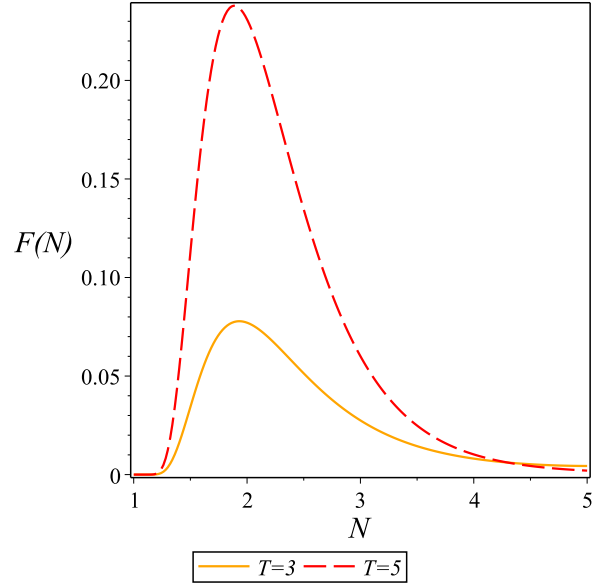}
 \end{array}$
 \end{center}
\caption{Boltzmann distribution function in terms of $N$ for the unit values of the parameters.}
 \label{fig6}
\end{figure}

The grand canonical partition function $Z_G$ can be defined as
\begin{equation}\label{eq4.1}
\ln{Z_{G}}=\beta PV.
\end{equation}
From chemical potential $\mu$ we find the fugacity $z$ as
\begin{equation}
\label{eq4.2}
z^N=e^{N\beta\mu}
\end{equation}
The distribution function follows as
\begin{equation}
\label{eq4.3}
F(N)=\frac{z^N{\cal Z}}{Z_G}
\end{equation}
which can be calculated as,
\[F(N)=\frac {\sqrt{2}\pi^{\frac {5} {2}}} {N^N N!}\left(\frac {m} {\beta N}\right)^{\frac {3} {2}}
\frac {k} {\beta N(N-1)Gm^2}
\phi\left(2,\frac {3} {2},-\alpha_0\right)\times\]
\begin{equation}
\label{eq4.4}
\exp\biggl{\{}-\left(\frac{7N-5}{2(N-1)}\right)-\frac{4(2N-1)}{3(N-1)}\phi\left(2,\frac {3} {2},-\alpha_0\right)-\frac {5N} {3}\left(1+\frac {8\alpha_0}{15}\frac{\phi\left(3,\frac {5} {2},-\alpha_0\right)}{\phi\left(2,\frac {3} {2},-\alpha_0\right)}\right)\biggr{\}}
\end{equation}
In Fig. \ref{fig6} we can see typical behavior of distribution function for the allowed values of $N$. We can see the most probable case is about $N=2$.
Using the above entropy, one can calculate specific heat via,

\setcounter{equation}{0}

\section{Tsallis partition function}

In this section, we have stepped to Tsallis approach for calculating the various important thermodynamic parameters, beginning from the partition function.
Tsallis q-exponential is defined as the distribution:
\begin{equation}
\label{eq5.1}
e_q(x)=[1+(q-1)x]_+^{\frac {1} {q-1}}
\end{equation}
Or equivalently
\begin{equation}
\label{ep5.2}
e_q(x)=
\begin{cases}
1+(q-1)x]^{\frac {1} {q-1}}\;\;\;;\;\;\;1+(q-1)x>0\\
0\;\;\;;\;\;\;1+(q-1)x<0
\end{cases}
\end{equation}
Let's consider the distribution
$\frac {1} {r}=PV\frac {1} {r}$.\\\\
Then $\frac {1} {r}\mid_{r=0}=0$.\\\\
We begin the calculation in $\nu$ dimensions starting from  $q>1$. Now
\begin{equation}
\label{eq5.3}
{\cal Z}_\nu=\int\limits_{-\infty}^{\infty}d^\nu x\int\limits_{-\infty}^{\infty}d^\nu p
\left[1+(q-1)\beta\left(\frac {N(N-1)Gm^2} {2r}+\frac{kN(N-1)Gm^2}{2r^2}-\frac {Np^2} {2m}\right)\right]_+^{\frac {1} {q-1}}
\end{equation}
Here, we put $q=4/3$ and get
\begin{eqnarray}
\label{eq5.4}
{\cal Z}_\nu&=&\left[\frac {2\pi^{\frac {\nu} {2}}} {\Gamma\left(\frac {\nu} {2}\right)}\right]^2
\int\limits_0^{\infty}p^{\nu-1} dp
\left[1+\beta\left(\frac {N(N-1)Gm^2} {6r}+\frac{N(N-1)Gm^2}{6r^2}-\frac {Np^2} {2m}\right)\right]_+^{3}\nonumber\\
&\times&
\int\limits_0^{\infty}r^{\nu-1}dr
\end{eqnarray}
After a series of calculations we get
\begin{eqnarray}
\label{eq5.5}
{\cal Z}_\nu&=&\left[\frac {2\pi^{\frac {\nu} {2}}} {\Gamma\left(\frac {\nu} {2}\right)}\right]^2\left(\frac{\beta N}{2m}\right)^3\left(\frac{(N-1)Gm^3}{3}\right)^{\frac{\nu}{2}+3}B(4,\frac{\nu}{2})\nonumber\\
&\times&\int\limits_{r_0}^{\infty}r^{-7}dr\left(\frac{6}{N(N-1)\beta Gm^2}r^2+r+k\right)^{\frac{\nu}{2}+3}
\end{eqnarray}
And we finally arrive at
\[{\cal Z}_\nu=\left[\frac {2\pi^{\frac {\nu} {2}}} {\Gamma\left(\frac {\nu} {2}\right)}\right]^2
\left(\frac{6m}{\beta N}\right)^{\frac {\nu} {2}} \frac {1}
{(3\alpha)\left[\frac {\alpha} {2}-\sqrt{\left(\frac {\alpha} {2}\right)^2-\alpha k}\right]^7}\left[\left(\frac {\alpha} {2}\right)^2-\alpha k\right]^{\frac {\nu+7} {2}}\]
\begin{equation}
\label{eq5.6}
\times{}_2F_1\left(7,4+\frac{\nu}{2};3-\frac{\nu}{2};z\right)B(4,\frac{\nu}{2})B(-\nu,4+\frac{\nu}{2})
\end{equation}
where
\begin{equation}
\label{eq5.7}
\alpha=\frac{N(N-1)\beta Gm^2}{6}
\end{equation}
and
\begin{equation}
\label{eq5.8}
z=\frac {\frac {\alpha} {2}+\sqrt{\left(\frac {\alpha} {2}\right)^2-\alpha k}}
{\frac {\alpha} {2}-\sqrt{\left(\frac {\alpha} {2}\right)^2-\alpha k}}.
\end{equation}
Using
\begin{equation}
\label{eq5.9}
{}_2F_1\left(7,4+\frac{\nu}{2};3-\frac{\nu}{2};z\right)=
(1-s)^7{}_2F_1\left(7,-1-\nu;3-\frac{\nu}{2};s\right),
\end{equation}
where
\begin{equation}
\label{eq5.10}
s=\frac {z} {z-1},
\end{equation}
and after a long calculation we obtain
\[\Gamma(-\nu){}_2F_1\left(7,4+\frac{\nu}{2};3-\frac{\nu}{2};z\right)=(1-s)^7\times\]
\begin{equation}
\label{eq5.11}
\left[\Gamma(-\nu)-(1+\nu)\sum\limits_{k=1}^4
\frac {\Gamma(k-\nu-1)(7)_k} {\left(3-\frac {\nu} {2}\right)}\frac {s^k} {k!}+\phi_\nu(s)\right],
\end{equation}
where $\phi_\nu(s)$ is regular in $\nu=3$. In particular:
\begin{equation}
\label{eq5.12}
\phi_3(s)=-\frac {4} {\Gamma(7)}\sum\limits_{k=0}^\infty
\frac {\Gamma(12+k)\Gamma(1+k)} {(11+2k)!!}
\frac{(2s)^{5+k}} {(5+k)!}
\end{equation}
Thus:
\begin{eqnarray}\label{eq5.13}
{\cal Z}_\nu&=&\left[\frac {2\pi^{\frac {\nu} {2}}} {\Gamma\left(\frac {\nu} {2}\right)}\right]^2
\left(\frac{6m}{\beta N}\right)^{\frac{\nu}{2}}
\frac{1}{(3\alpha)\left[\frac {\alpha} {2}-\sqrt{\left(\frac {\alpha} {2}\right)^2-\alpha k}\right]^7}
\left[\left(\frac {\alpha} {2}\right)^2-\alpha k\right]^{\frac {\nu+7} {2}}(1-s)^7\nonumber\\
&\times&\frac{6\Gamma\left(\frac {\nu} {2}\right)} {\Gamma\left(4-\frac {\nu} {2}\right)}\left[\Gamma(-\nu)-(1+\nu)\sum\limits_{k=1}^4
\frac{\Gamma(k-\nu-1)(7)_k} {\left(3-\frac {\nu} {2}\right)}\frac {s^k} {k!}+\phi_\nu(s)\right].
\end{eqnarray}
Now, we do the Laurent development explicitly, taking into account that $\phi_\nu(s)$ does not have a pole at $\nu = 3$ and take the expression for $\phi_3$ at the end of the development.
Thus, the Laurent development goes as
\begin{equation}
\label{eq5.14}
{\cal Z}_\nu=f(\nu)\Gamma(3-\nu)+g(\nu)\phi_\nu(s).
\end{equation}
Here,
\[g(\nu)=\left[\frac {192\pi^{\nu-1}\sin\frac{\nu\pi}{2}}{(6-\nu)(2-\nu)(4-\nu)}\right]\left(\frac{6m}{\beta N}\right)^{\frac {\nu} {2}}\frac {\left[\left(\frac{\alpha}{2}\right)^2-\alpha k\right]^{\frac {\nu+7} {2}}}
{(3\alpha)\left[\frac {\alpha} {2}-\sqrt{\left(\frac {\alpha} {2}\right)^2-\alpha k}\right]^7}
(1-s)^7\]
and
\begin{eqnarray}\label{eq5.15}
f(\nu)\left(\Gamma(3-\nu)\right)&=&\left[\frac {192\pi^{\nu-1}\sin\frac{\nu\pi}{2}}{(6-\nu)(-\nu)(1-\nu)(2-\nu)^2(4-\nu)}\right]\left(\frac{6m}{\beta N}\right)^{\frac{\nu}{2}}\nonumber\\
&\times&(1-s)^7\left(1-14s\frac{(1+\nu)}{(6-\nu)}X\right)\left(\Gamma\left(3-\nu\right)\right)\nonumber\\
&\times&\frac{\left[\left(\frac{\alpha}{2}\right)^2-\alpha k\right]^{\frac {\nu+7} {2}}}
{(3\alpha)\left[\frac {\alpha} {2}-\sqrt{\left(\frac {\alpha} {2}\right)^2-\alpha k}\right]^7},
\end{eqnarray}
where
\begin{eqnarray}\label{eq5.16}
f(\nu)&=&\left[\frac{192\pi^{\nu-1}\sin\frac{\nu\pi}{2}}{(6-\nu)(-\nu)(1-\nu)(2-\nu)^2(4-\nu)}\right]\left(\frac{6m}{\beta N}\right)^{\frac{\nu}{2}}\nonumber\\
&\times&\frac{\left[\left(\frac{\alpha}{2}\right)^2-\alpha k\right]^{\frac{\nu+7}{2}}}
{(3\alpha)\left[\frac {\alpha} {2}-\sqrt{\left(\frac{\alpha}{2}\right)^2-\alpha k}\right]^{7}}(1-s)^{7}(1-14s\frac{(1+\nu)}{(6-\nu)}X),
\end{eqnarray}
and
\begin{eqnarray}
X&=&1+\frac{4s(-\nu)}{4-\frac{\nu}{2}}+\frac{12s^2(-\nu)
(1-\nu)}{(4-\frac{\nu}{2})(5-\frac{\nu}{2})}\nonumber\\
&+&\frac{30s^3(-\nu)(1-\nu)(2-\nu)}{(4-\frac{\nu}{2})(5-\frac{\nu}{2})(6-\frac{\nu}{2})}
\end{eqnarray}
Therefore,
\begin{eqnarray}\label{eq5.17}
f(3)&=&\frac{32\pi^2}{9}\left(\frac{6m}{\beta N}\right)^{\frac {3} {2}}
(3\alpha)^{-1}\left[\left(\frac {\alpha} {2}\right)^2-\alpha k\right]^{5}(1-s)^7\nonumber\\
&\times&\left[3-56s\left(1-\frac{24s}{5}+\frac{288s^2}{35}-\frac{32s^3}{7}\right)\right].
\end{eqnarray}
Also, we can obtain,
\begin{eqnarray}\label{eq5.18}
\frac{f^{'}(\nu)}{f(\nu)}&=&\frac{-1-14s(1-12s^2(1-3\nu^2)-30s^3(1-4\nu-6\nu^2+4\nu^3)-4s(1+2\nu))}
{\biggl(6-\nu-14s(1+\nu)\left[1-4s(\nu)-12s^2(\nu)(1-\nu)-30s^3(\nu)(1-\nu)(2-\nu)\right]\biggr)}\nonumber\\
&+&\frac{1}{2}\ln\left(\pi^2\frac{6m}{\beta N}\left[\left(\frac{\alpha}{2}\right)^2-\alpha k\right]\right)\nonumber\\
&+&\frac{\pi}{2}\cot\frac{\nu\pi}{2}+\frac{2}{6-\nu}-\frac{1}{\nu}+\frac{1}{1-\nu}+\frac{2}{2-\nu}+\frac{1}{4-\nu}
\end{eqnarray}
which yields to the following relation,
\[\frac{f^{'}(3)}{f(3)}=\frac{1}{2}\ln\left(\pi^2\frac{6m}{\beta N}\left[\left(\frac{\alpha}{2}\right)^2-\alpha k\right]\right)-\frac{7}{6}\]
\begin{equation}
\label{eq5.19}
-\left[\frac {14s\left(-\frac{1}{3}-\frac{536s}{75}+\frac{70944s^2}{3675}-\frac{347232s^3}{13545})\right)}{\biggl(3-56s\left(1-\frac{24s}{5}+\frac{288s^2}{35}-\frac{32s^3}{7}\right)\biggr)}\right]
\end{equation}
Thus
\begin{equation}
\label{eq5.20}
f(\nu)=f(3)+f^{'}(3)(\nu-3)+\sum\limits_{k=2}^\infty b_k(\nu-3)^k
\end{equation}
For the $\Gamma$ function we have
\begin{equation}
\label{eq5.21}
\Gamma(3-\nu)=\frac {1} {3-\nu}-C+\sum\limits_{k=1}^\infty c_k\left(\nu-3\right)^k
\end{equation}
As a consequence:
\begin{equation}
\label{eq5.22}
{\cal Z}_\nu=\frac {f(3)} {3-\nu}-f(3)C-f^{'}(3)+\sum\limits_{k=1}^{\infty}a_k\left(\nu-3\right)^k+g(3)\phi_3(s)
\end{equation}
and
\begin{equation}
\label{eq5.23}
{\cal Z}=-f(3)\left(C+\frac{f^{'}(3)}{f(3)}\right)+g(3)\phi_3(s)
\end{equation}
Since
\begin{equation}
\label{eq5.24}
g(3)=-64\left(\frac{6m}{\beta N}\right)^{\frac {3} {2}}
\frac{\left[\left(\frac {\alpha} {2}\right)^2-\alpha k\right]^{5}}
{(3\alpha)\left[\frac {\alpha} {2}-\sqrt{\left(\frac {\alpha} {2}\right)^2-\alpha k}\right]^7}
(1-s)^7
\end{equation}
therefore
\begin{equation}
\label{eq5.25}
f(3)=\frac{1}{18}g(3)\left[3-56s\left(1-\frac{24s}{5}+\frac{288s^2}{35}-\frac{32s^3}{7}\right)\right]
\end{equation}
where $C$ is the Euler-Mascheroni constant.
As a consequence:
\begin{equation}
\label{eq5.26}
{\cal Z}=\frac{1}{18}g(3)\left[ \left[3-56s\left(1-\frac{24s}{5}+\frac{288s^2}{35}-\frac{32s^3}{7}\right)\right]\left(C+\frac{f^{'}(3)}{f(3)}\right)+18\phi_3(s)\right]
\end{equation}
Or equivalently:
\[{\cal Z}=-\frac{32}{9}\left(\frac{6m}{\beta N}\right)^{\frac {3} {2}}
\frac{\left(\left(\frac {\alpha} {2}\right)^2-\alpha k\right)^{5}}
{(3\alpha)\left[\frac {\alpha} {2}-\sqrt{\left(\frac {\alpha} {2}\right)^2-\alpha k}\right]^7}
(1-s)^7\]
\[\biggl[\left[3-56s\left(1-\frac{24s}{5}+\frac{288s^2}{35}-\frac{32s^3}{7}\right)\right]\biggl(C+\frac{1}{2}\ln\left(\pi^2\frac{6m}{\beta N}\left(\left(\frac{\alpha}{2}\right)^2-\alpha k\right)\right)\]
\begin{equation}
\label{eq5.27}
-\frac{7}{6}-\left[\frac {14s\left(-\frac{1}{3}-\frac{536s}{75}+\frac{70944s^2}{3675}-\frac{347232s^3}{13545})\right)}{\biggl(3-56s\left(1-\frac{24s}{5}+\frac{288s^2}{35}-\frac{32s^3}{7}\right)\biggr)}\right]+18\phi_3(s)\biggr]
\end{equation}
We find that partition function is completely a positive quantity, which is decreasing function of temperature, and is of  Gaussian nature.

\subsection{Mean energy}

In this section, Tsallis approach has further been extended to relate partition function $\cal Z $ and internal energy $\cal U$ as
\[{\cal Z}<{\cal U}>_\nu=\left[\frac {2\pi^{\frac {\nu} {2}}} {\Gamma\left(\frac {\nu} {2}\right)}\right]^2
\int\limits_0^{\infty}r^{\nu-1}dr
\int\limits_0^{P_0} p^{\nu-1} dp
\left(-\frac {N(N-1)Gm^2} {2r}-\frac {kN(N-1)Gm^2} {2r^2}+\frac {Np^2} {2m}\right)\]
\begin{equation}
\label{eq6.1}
\times\left[1+(q-1)\beta\left(\frac {N(N-1)Gm^2} {2r}+\frac{kN(N-1)Gm^2}{2r^2}-\frac{Np^2}{2m}\right)\right]^{\frac {1} {q-1}}
\end{equation}
This can be rearranged as
\[{\cal Z}<{\cal U}>_\nu=\biggl[\frac{2\pi^\frac{\nu}{2}}{\Gamma{\frac{\nu}{2}}}\biggr]^2[\frac{\beta N(q-1)}{2m}]^{\frac{1}{q-1}}\]
\[\left(\int\limits_0^{\infty}r^{\nu-1}dr
\int\limits_0^{P_0}\frac{Np^{\nu+1}}{2m}\left[\frac{1}{\beta N(q-1)}+\frac{(N-1)Gm^2}{2r}+\frac{k(N-1)Gm^2}{2r^2}-\frac{p^2}{2m}\right]^{\frac{1}{q-1}}dp\right.\]
\[\left.-\frac {N(N-1)Gm^2} {2}\int\limits_0^{\infty}r^{\nu-2}dr\int\limits_0^{P_0}p^{\nu-1}\left[\frac{1}{\beta N(q-1)}+\frac{(N-1)Gm^2}{2r}+\frac{k(N-1)Gm^2}{2r^2}-\frac{p^2}{2m}\right]^{\frac{1}{q-1}}dp\right)\]
\begin{equation}
\label{eq6.2}
\left.-\frac {kN(N-1)Gm^2} {2}\int\limits_0^{\infty}r^{\nu-3}dr\int\limits_0^{P_0}p^{\nu-1}\left[\frac{1}{\beta N(q-1)}+\frac{(N-1)Gm^2}{2r}+\frac{k(N-1)Gm^2}{2r^2}-\frac{p^2}{2m}\right]^{\frac{1}{q-1}}dp\right)
\end{equation}
For $q=4/3$, we get
\[{\cal Z}<U>_\nu=\frac {3} {2\beta}\left[\frac {2\pi^{\frac {\nu} {2}}} {\Gamma\left(\frac {\nu} {2}\right)}\right]^2
\left(\frac {6m} {N\beta}\right)^{\frac {\nu} {2}}
\frac {\left[\left(\frac {\alpha} {2}\right)^2-\alpha k\right]^{\frac {\nu+7} {2}}}
{\left[\frac {\alpha} {2}-\sqrt{\left(\frac {\alpha} {2}\right)^2-\alpha k}\right]^9}\times \]
\[\biggl{\{}B\left(4,\frac {\nu} {2}+1\right)B\left(-\nu,\frac {\nu} {2}+5\right)
{}_2F_1\left(9,\frac {\nu} {2}+5,4-\frac {\nu} {2},z\right)\]\[
-\alpha\left[\frac {\alpha} {2}-\sqrt{\left(\frac {\alpha} {2}\right)^2-\alpha k}\right]B\left(4,\frac {\nu} {2}\right)B\left(1-\nu,\frac {\nu} {2}+4\right)
{}_2F_1\left(8,\frac {\nu} {2}+4,4-\frac {\nu} {2},z\right)\]
\begin{equation}
\label{eq6.3}
-k\alpha B\left(4,\frac {\nu} {2}\right)B\left(2-\nu,\frac {\nu} {2}+4\right)
{}_2F_1\left(9,\frac {\nu} {2}+4,5-\frac {\nu} {2},z\right)\biggr{\}}
\end{equation}
which reduces to
\[{\cal Z}<U>_\nu=\frac {3\nu} {4\beta}\frac{6\Gamma\left(\frac{\nu}{2}\right)}{\Gamma\left(5-\frac{\nu}{2}\right)}\left[\frac {2\pi^{\frac {\nu} {2}}} {\Gamma\left(\frac {\nu} {2}\right)}\right]^2
\left(\frac {6m} {N\beta}\right)^{\frac {\nu} {2}}
\frac {\left[\left(\frac {\alpha} {2}\right)^2-\alpha k\right]^{\frac {\nu+7} {2}}}
{\left[\frac {\alpha} {2}-\sqrt{\left(\frac {\alpha} {2}\right)^2-\alpha k}\right]^9}\times \]
\[\biggl{\{}\Gamma(-\nu)
{}_2F_1\left(9,\frac {\nu} {2}+5,4-\frac {\nu} {2},z\right)\]\[
+\left({\alpha}^2-\alpha\sqrt{{\alpha}^2-4\alpha k}\right)\Gamma(-\nu)
{}_2F_1\left(8,\frac {\nu} {2}+4,4-\frac {\nu} {2},z\right)\]
\begin{equation}
\label{eq6.4}
+2k\alpha \frac{(1-\nu)}{\left(10-\nu\right)}\Gamma(-\nu)
{}_2F_1\left(9,\frac {\nu} {2}+4,5-\frac {\nu} {2},z\right)\biggr{\}}
\end{equation}
Now, using the transformation formula
\begin{equation}
\label{eq6.5}
{}_2F_1\left(a,b,c,z\right)=(1-z)^{-a}{}_2F_1\left(a,c-b,c,\frac{z}{z-1}\right)
\end{equation}
we get\\\\
${}_2F_1\left(9,\frac {\nu} {2}+5,4-\frac {\nu} {2},z\right)=(1-s)^9{}_2F_1\left(9,-1-\nu,4-\frac {\nu} {2},s\right)$\\
${}_2F_1\left(8,\frac {\nu} {2}+4,4-\frac {\nu} {2},z\right)=(1-s)^8{}_2F_1\left(8,-\nu,4-\frac {\nu} {2},s\right)$\\
${}_2F_1\left(9,\frac {\nu} {2}+4,5-\frac {\nu} {2},z\right)=(1-s)^9{}_2F_1\left(9,1-\nu,5-\frac {\nu} {2},s\right)$\\\\
Consequently equation (\ref{eq6.4}) gets modified as
\[{\cal Z}<U>_\nu=-\frac {288\pi^{\nu-1}\nu\sin(\frac{\nu\pi}{2})} {\beta(\nu-2)(\nu-4)(\nu-6)(\nu-8)}
\left(\frac {6m} {N\beta}\right)^{\frac {\nu} {2}}
\frac {\left[\left(\frac {\alpha} {2}\right)^2-\alpha k\right]^{\frac {\nu+7} {2}}}
{\left[\frac {\alpha} {2}-\sqrt{\left(\frac {\alpha} {2}\right)^2-\alpha k}\right]^9}\times \]
\[\biggl{\{}(1-s)^9\Gamma(-\nu){}_2F_1\left(9,-1-\nu,4-\frac {\nu} {2},s\right)\]\[
+\left({\alpha}^2-\alpha\sqrt{{\alpha}^2-4\alpha k}\right)
(1-s)^8\Gamma(-\nu){}_2F_1\left(8,-\nu,4-\frac {\nu} {2},s\right)\]
\begin{equation}
\label{eq6.6}
+2k\alpha \frac{(1-\nu)}{\left(10-\nu\right)}
(1-s)^9\Gamma(-\nu){}_2F_1\left(9,1-\nu,5-\frac {\nu} {2},s\right) \biggr{\}}
\end{equation}
which can further be rearranged as
\[{\cal Z}<U>_\nu=g(\nu)\biggl{\{}(1-s)\left(h_1(\nu)\Gamma(3-\nu)+\phi_1(s,\nu)\right)+\]
\begin{equation}
\label{eq6.7}
\left({\alpha}^2-\alpha\sqrt{{\alpha}^2-4\alpha k}\right)\left(h_2(\nu)\Gamma(3-\nu)+\phi_2(s,\nu)
\right)+\frac{2k\alpha(1+s)}{10-\nu}\left(h_3(\nu)\Gamma(3-\nu)+\phi_3(s,\nu)\right)\biggr{\}}
\end{equation}
We now use the following equalities:
\[\Gamma(-\nu){}_2F_1\left(9,-1-\nu,4-\frac {\nu} {2},s\right)=\left(h_1(\nu)\Gamma(3-\nu)+\phi_1(s,\nu)\right)\]
\[\Gamma(-\nu){}_2F_1\left(8,-\nu,4-\frac {\nu} {2},s\right)=\left(h_2(\nu)\Gamma(3-\nu)+\phi_2(s,\nu)\right)\]
\[\Gamma(-\nu){}_2F_1\left(9,1-\nu,5-\frac {\nu} {2},s\right)=\left(h_3(\nu)\Gamma(3-\nu)+\phi_3(s,\nu)\right)\]
where
\begin{equation}
\label{eq6.8}
g(\nu)=-\frac {288\pi^{\nu-1}\nu\sin(\frac{\nu\pi}{2})} {\beta(\nu-2)(\nu-4)(\nu-6)(\nu-8)}
\left(\frac {6m} {N\beta}\right)^{\frac {\nu} {2}}
\frac {\left[\left(\frac {\alpha} {2}\right)^2-\alpha k\right]^{\frac {\nu+7} {2}}}
{\left[\frac {\alpha} {2}-\sqrt{\left(\frac {\alpha} {2}\right)^2-\alpha k}\right]^9}(1-s)^8
\end{equation}
In three dimensions we have:
\begin{equation}
\label{eq6.9}
g(3)=-\frac {288\pi^{2}} {5\beta}
\left(\frac {6m} {N\beta}\right)^{\frac {3} {2}}
\frac {\left[\left(\frac {\alpha} {2}\right)^2-\alpha k\right]^{5}}
{\left[\frac {\alpha} {2}-\sqrt{\left(\frac {\alpha} {2}\right)^2-\alpha k}\right]^9}(1-s)^8
\end{equation}
And for its derivative:
\begin{equation}
\label{eq6.10}
g^{'}(3)=g(3)\left[\frac{13}{15}+\frac{1}{2}\ln\left(\pi^2\frac{Gm}{N\beta}\left(\frac{\alpha^2}{4}-\alpha k\right)\right)\right]
\end{equation}
We now make the following definitions:
\[h_1(\nu)=\biggl[\frac{1}{(-\nu)(1-\nu)(2-\nu)}-(1+\nu)\biggl(\frac{9s}{(-\nu)(1-\nu)(2-\nu)(4-\frac{\nu}{2})}+\]\[\frac{9.10s^2}{2(1-\nu)(2-\nu)(4-\frac{\nu}{2})(5-\frac{\nu}{2})}+\frac{9.10.11s^3}{3!(2-\nu)(4-\frac{\nu}{2})(5-\frac{\nu}{2})(6-\frac{\nu}{2})}\]
\begin{equation}
\label{eq6.11}
+\frac{9.10.11.12s^4}{4!(4-\frac{\nu}{2})(5-\frac{\nu}{2})(6-\frac{\nu}{2})(7-\frac{\nu}{2})}\biggr)\biggr]
\end{equation}
In three dimensions:
\begin{equation}
\label{eq6.12}
h_1(3)=\left[-\frac{1}{6}+\frac{36s}{15}-\frac{72s^2}{7}+\frac{352s^3}{21}-\frac{64s^4}{7}\right]
\end{equation}
The second function $h_2$ is given by:
\[h_2(\nu)=\biggl[\frac{1}{(-\nu)(1-\nu)(2-\nu)}+\frac{8s}{(\nu)(1-\nu)(2-\nu)(4-\frac{\nu}{2})}+\]
\begin{equation}
\label{eq6.13}
\frac{8.9s^2}{2(2-\nu)(4-\frac{\nu}{2})(5-\frac{\nu}{2})}+\frac{8.9.10s^3}{3!(4-\frac{\nu}{2})(5-\frac{\nu}{2})(6-\frac{\nu}{2})}\biggr]
\end{equation}
In three dimensions:
\begin{equation}
\label{eq6.14}
h_2(3)=\left[-\frac{1}{6}+\frac{8s}{15}-\frac{144s^2}{35}+\frac{64s^3}{21}\right]
\end{equation}
The third function is defined as:
\[h_3(\nu)=\biggl[\frac{1}{(-\nu)(1-\nu)(2-\nu)}+\frac{9s}{(-\nu)(2-\nu)(5-\frac{\nu}{2})}+\frac{9.10s^2}{2(-\nu)(5-\frac{\nu}{2})(6-\frac{\nu}{2})}\biggr]\]
And in three dimensions:
\begin{equation}
\label{eq6.15}
h_3(3)=\left[-\frac{1}{6}+\frac{6s}{7}-\frac{20s^2}{21}\right]
\end{equation}
We also define:
\[\phi_1(s,\nu)=\biggl[(-1-\nu)\frac{9.10.11.12.13\Gamma\left(4-\nu\right)s^5}{5!(4-\frac{\nu}{2})(5-\frac{\nu}{2})(6-\frac{\nu}{2})(7-\frac{\nu}{2})}+\]
\begin{equation}
\label{eq6.16}
(-1-\nu)\frac{9.10.11.12.13.14\Gamma\left(5-\nu\right)s^6}{6!(4-\frac{\nu}{2})(5-\frac{\nu}{2})(6-\frac{\nu}{2})(7-\frac{\nu}{2})(8-\frac{\nu}{2})}+....\biggr]
\end{equation}
\begin{equation}
\label{eq6.17}
\phi_2(s,\nu)=\biggl[\frac{8.9.10.11\Gamma\left(4-\nu\right)s^4}{4!(4-\frac{\nu}{2})(5-\frac{\nu}{2})(6-\frac{\nu}{2})(7-\frac{\nu}{2})}+\frac{8.9.10.11.12\Gamma\left(5-\nu\right)s^5}{5!(4-\frac{\nu}{2})(5-\frac{\nu}{2})(6-\frac{\nu}{2})(7-\frac{\nu}{2})(8-\frac{\nu}{2})}+....\biggr]
\end{equation}
\begin{equation}
\label{eq6.18}
\phi_3(s,\nu)=\biggl[\frac{9.10.11\Gamma\left(4-\nu\right)s^3}{3!(-\nu)(5-\frac{\nu}{2})(6-\frac{\nu}{2})(7-\frac{\nu}{2})}+\frac{9.10.11.12\Gamma\left(5-\nu\right)s^4}{4!(-\nu)(5-\frac{\nu}{2})(6-\frac{\nu}{2})(7-\frac{\nu}{2})(8-\frac{\nu}{2})}+....\biggr]
\end{equation}
Using this functions we can write:
\[{\cal Z}<U>_\nu=g(\nu)\biggl{\{}f_1(\nu)\Gamma(3-\nu)+\Phi_1(s,\nu)+f_2(\nu)\Gamma(3-\nu)+\Phi_2(s,\nu)+f_3(\nu)\Gamma(3-\nu)+\Phi_3(s,\nu)\biggr{\}}=\]
\begin{equation}
\label{eq6.19}
g(\nu)\left\{\Gamma(3-\nu)\left(f_1(\nu)+f_2(\nu)+f_3(\nu)\right)+\Phi_2(s,\nu)+\Phi_1(s,\nu)+\Phi_3(s,\nu)\right\}=
\end{equation}
\begin{equation}
\label{eq6.20}
g(\nu)\biggl{\{}\Gamma(3-\nu)h(\nu)+\phi_\nu(s)\biggr{\}}
\end{equation}
where
\begin{equation}
\label{eq6.21}
h(\nu)=\left(f_1(\nu)+f_2(\nu)+f_3(\nu)\right)
\end{equation}
\begin{equation}
\label{eq6.22}
\Phi_\nu(s)=\Phi_2(s,\nu)+\Phi_1(s,\nu)+\Phi_3(s,\nu)
\end{equation}
\begin{equation}
\label{eq6.23}
f_1(\nu)=(1-s)h_1(\nu)
\end{equation}
\begin{equation}
\label{eq6.24}
f_2(\nu)=\left({\alpha}^2-\alpha\sqrt{{\alpha}^2-4\alpha k}\right)h_2(\nu)
\end{equation}
\begin{equation}
\label{eq6.25}
f_3(\nu)=\frac{2k\alpha(1-s)}{10-\nu}h_3(\nu)
\end{equation}
\begin{equation}
\label{eq6.26}
\Phi_1(s,\nu)=(1-s)\phi_1(s,\nu)
\end{equation}
\begin{equation}
\label{eq6.27}
\Phi_2(s,\nu)=\left({\alpha}^2-\alpha\sqrt{{\alpha}^2-4\alpha k}\right)\phi_2(s,\nu)
\end{equation}
\begin{equation}
\label{eq6.28}
\Phi_3(s,\nu)=\frac{2k\alpha(1-s)}{10-\nu}\phi_3(s,\nu)
\end{equation}
For the Laurent's development we go as
\begin{equation}
\label{eq6.29}
\left(g(\nu)h(\nu)\right)^{'}=g(\nu)\left[f_1^{'}(\nu)+f_2^{'}(\nu)+f_3^{'}(\nu)+\frac{g^{'}(\nu)}{g(\nu)}\left(f_1(\nu)+f_2(\nu)+f_2(\nu)\right)\right]
\end{equation}
Thus
\[\left(g(3)f(3)\right)^{'}=g(3)\left[f_1^{'}(3)+f_2^{'}(3)+f_3^{'}(3)+\frac{g^{'}(3)}{g(3)}\left(f_1(3)+f_2(3)+f_2(3)\right)\right]\]
\[f_1^{'}(3)=(1-s)\left[\frac{11}{36}-\frac{63s}{25}+\frac{2286s^2}{245}-\frac{32824s^3}{6615}-\frac{176272s^4}{24255}\right]\]
\[f_2^{'}(3)=\left({\alpha}^2-\alpha\sqrt{{\alpha}^2-4\alpha k}\right)\left[\frac{11}{36}-\frac{196s}{225}+\frac{3312s^2}{1225}+\frac{9152s^3}{6615}\right]\]
\[f_3^{'}(3)={2k\alpha(1-s)}\left[\frac{71}{1764}-\frac{63s}{539}-\frac{80s^2}{9261}\right]\]

\begin{equation}
\label{eq6.30}
g(\nu)h(\nu)=g(3)f(3)+(g(3)f(3))^{'}(\nu-3)+\sum\limits_{k=2}^\infty b_k(\nu-3)^k
\end{equation}
For the $\Gamma$ function we have
\begin{equation}
\label{eq6.31}
\Gamma(3-\nu)=\frac {1} {3-\nu}-C+\sum\limits_{k=1}^\infty c_k\left(\nu-3\right)^k
\end{equation}As a consequence:
\begin{equation}
\label{eq6.32}
{\cal Z}<U>_\nu=\frac {g(3)h(3)} {3-\nu}-g(3)h(3)C-(g(3)h(3))^{'}+\sum\limits_{k=1}^{\infty}a_k\left(\nu-3\right)^k+g(3)\Phi(3,s)
\end{equation}
and
\begin{equation}
\label{eq6.33}
{\cal Z}<U>=-g(3)h(3)\left(C+\frac{(g(3)h(3))^{'}}{g(3)h(3)}\right)+g(3)\Phi(3,s)
\end{equation}
Therefore, we can obtain,
\begin{equation}
\label{eq6.34}
<U>= -\frac{h(3)\left(C+\frac{(g(3)h(3))^{'}}{g(3)h(3)}\right)+\Phi(3,s)}{\left[ \left[\frac{1}{6}-\frac{28}{9}s\left(1-\frac{24s}{5}+\frac{288s^2}{35}-\frac{32s^3}{7}\right)\right]\left(C+\frac{f^{'}(3)}{f(3)}\right)+\phi_3(s)\right]}
\end{equation}

\subsection{Stability}

\begin{figure}[h!]
 \begin{center}$
 \begin{array}{cccc}
\includegraphics[width=70 mm]{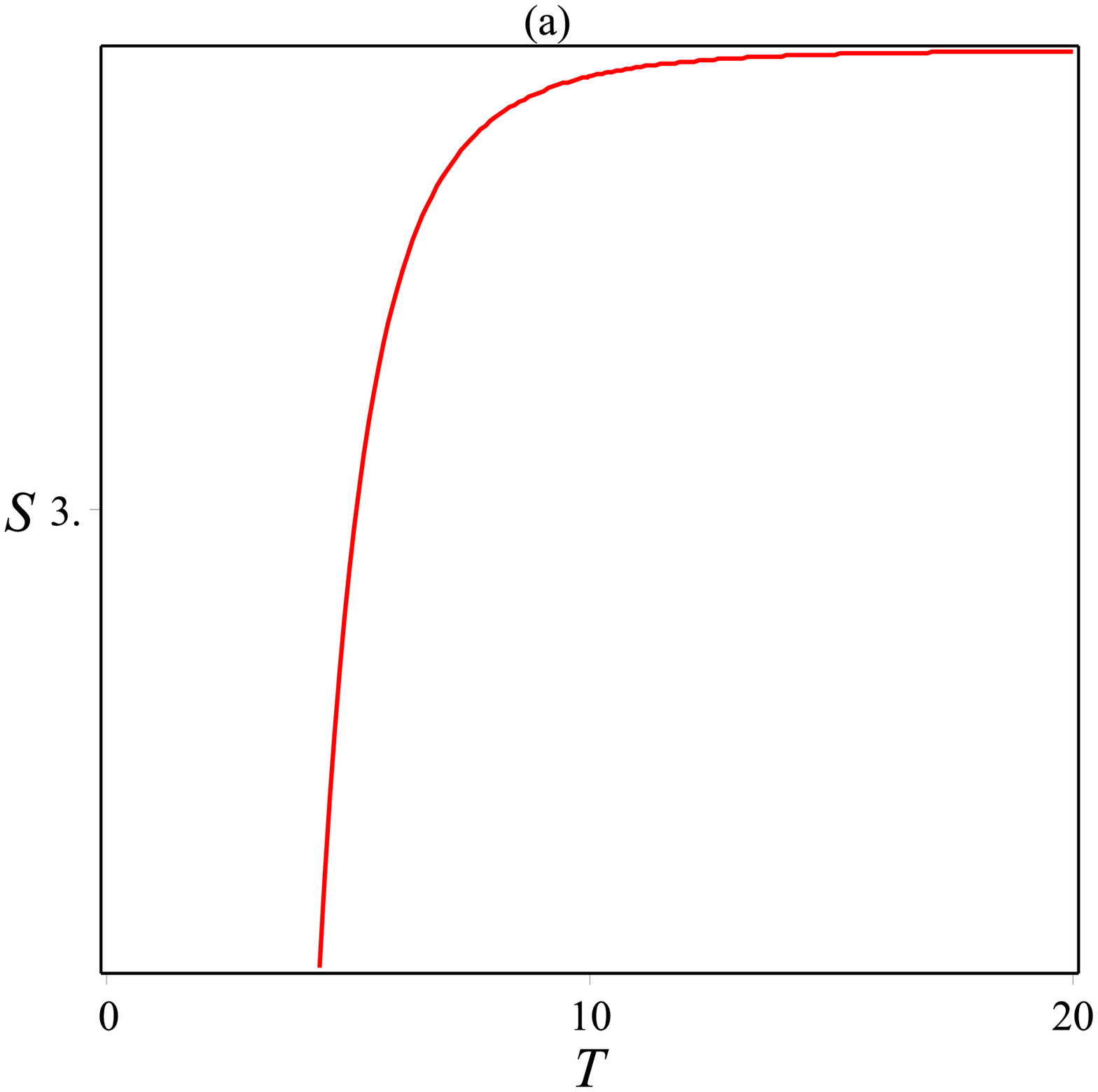}\includegraphics[width=70 mm]{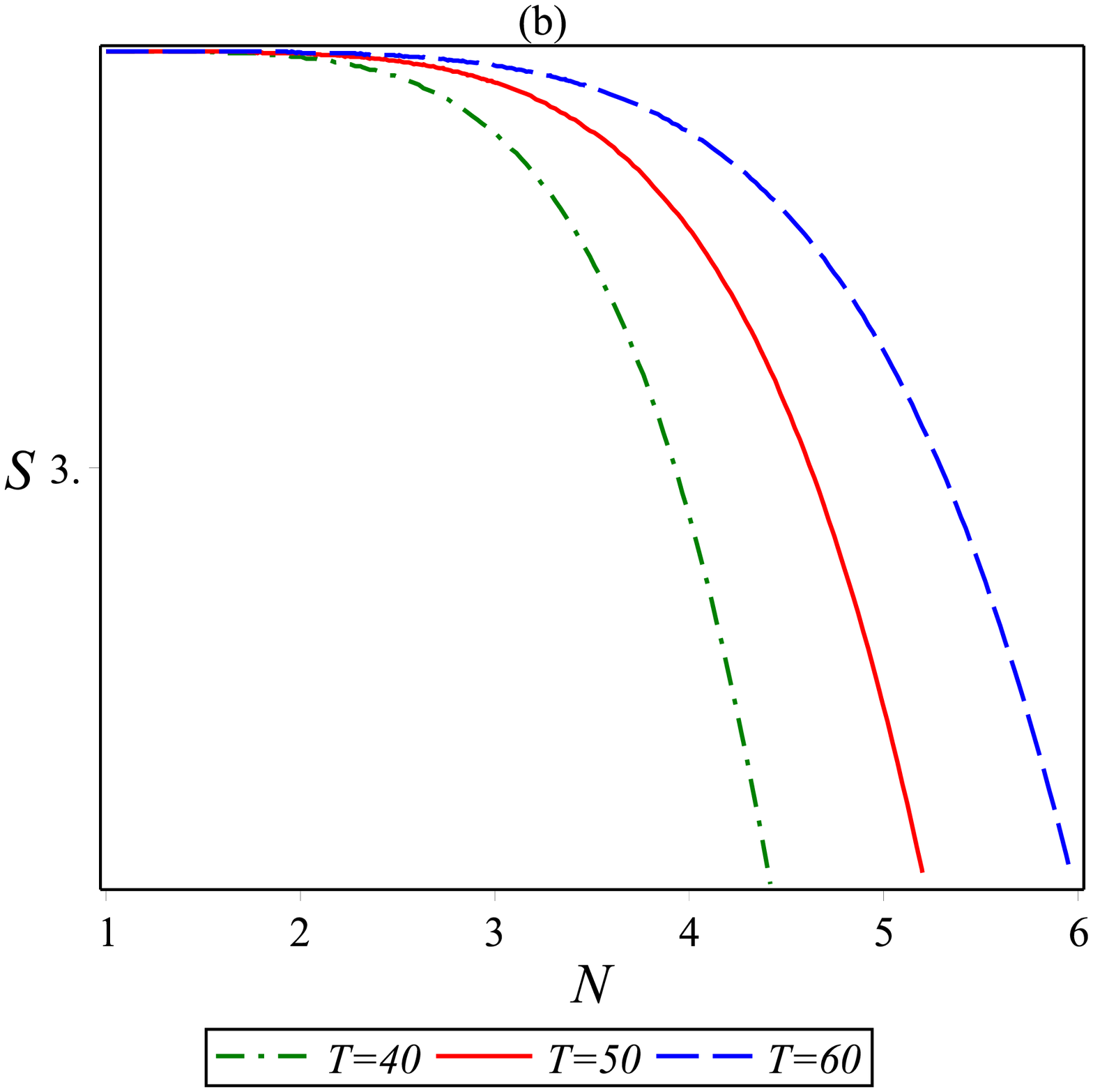}
 \end{array}$
 \end{center}
\caption{Typical behavior of the entropy in terms of (a) $T$, and (b) $N$ for the unit values of the parameters.}
 \label{figS2}
\end{figure}

In this subsection, we are going to investigate the stability of this model by analyzing the specific heat. We begin it by evaluating the entropy ${\cal S}$ using the formula
\begin{equation}
\label{eq7.1}
{\cal S}=\ln_{\frac {4} {3}}{\cal Z}+{\cal Z}^{-\frac {1} {3}}\beta<{\cal U}>=3+(\beta<{\cal U}>-3){\cal Z}^{-\frac{1}{3}},
\end{equation}
which yields the following expression,
\[{\cal S}=3-\left(\beta\frac{h(3)\left(C+\frac{(g(3)h(3))^{'}}{g(3)h(3)}\right)+\Phi(3,s)}{\left[ \left[\frac{1}{6}-\frac{28}{9}s\left(1-\frac{24s}{5}+\frac{288s^2}{35}-\frac{32s^3}{7}\right)\right]\left(C+\frac{f^{'}(3)}{f(3)}\right)+\phi_3(s)\right]}+3\right)\]
\begin{equation}
\label{eq7.4}
  {\biggl(\frac{1}{18}g(3)\left[ \left[3-56s\left(1-\frac{24s}{5}+\frac{288s^2}{35}-\frac{32s^3}{7}\right)\right]\left(C+\frac{f^{'}(3)}{f(3)}\right)+18\phi_3(s)\right]\biggr)}^{-\frac{1}{3}}
\end{equation}
We have drawn the entropy in Fig. \ref{figS2} to see the typical behavior in terms of $T$. We can see that it is an increasing function of the temperature and yields to a constant at high temperature (see Fig. \ref{figS2} (a)) thus verifying the second law of thermodynamics. However, we find a minimum temperature for the given system below which the entropy is not well defined and becomes negative. Also, in Fig. \ref{figS2} (b) we  have drawn entropy in terms of $N$ and it  comes out to be a decreasing function of $N$. It is observed that there is an upper limit for the number of galaxies, and it is similar to the previous case of Boltzmann statistics. We will show that this upper limit is constrained further by using stability analysis of specific heat.\\

\begin{figure}[h!]
 \begin{center}$
 \begin{array}{cccc}
\includegraphics[width=70 mm]{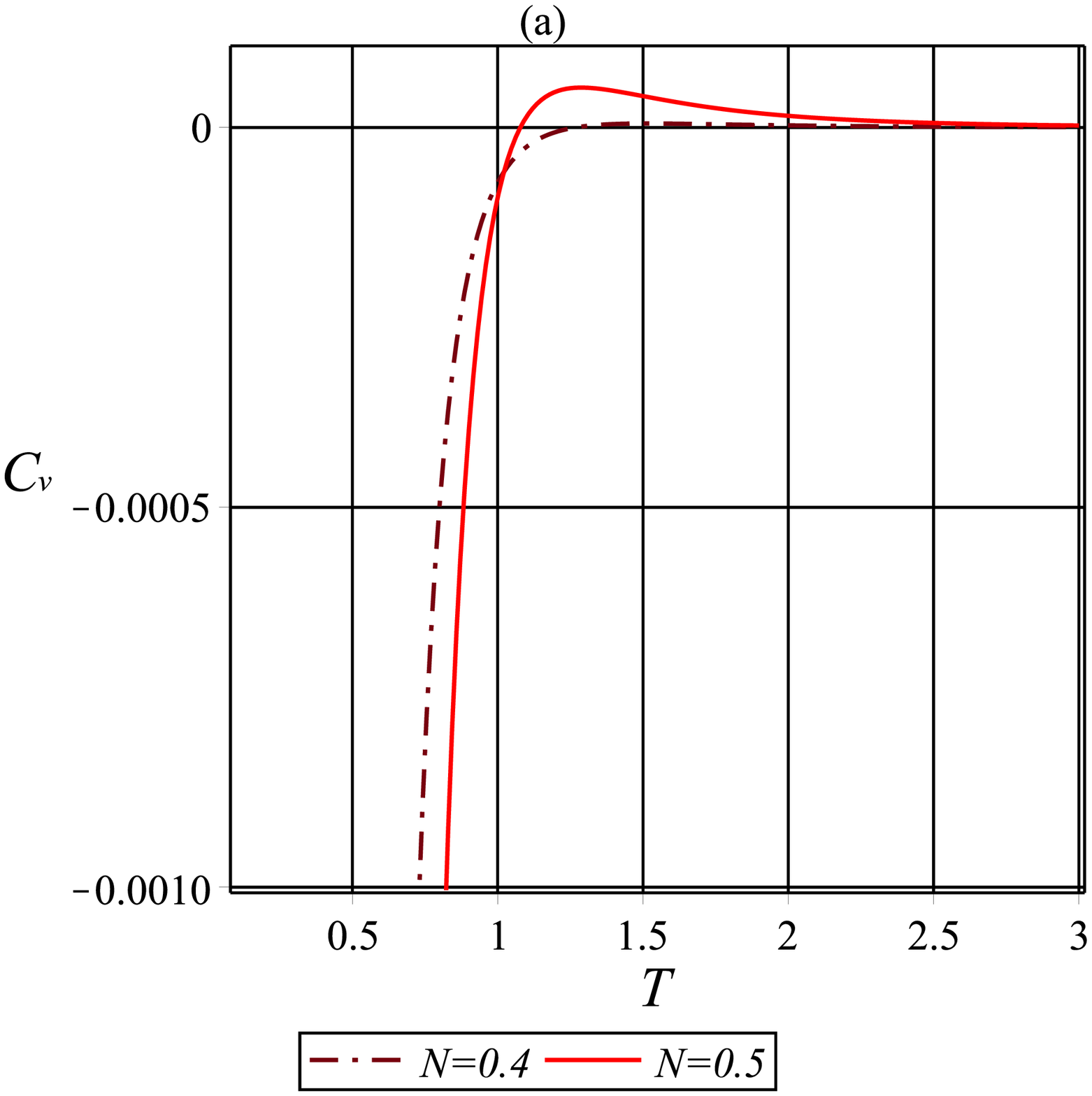}\includegraphics[width=70 mm]{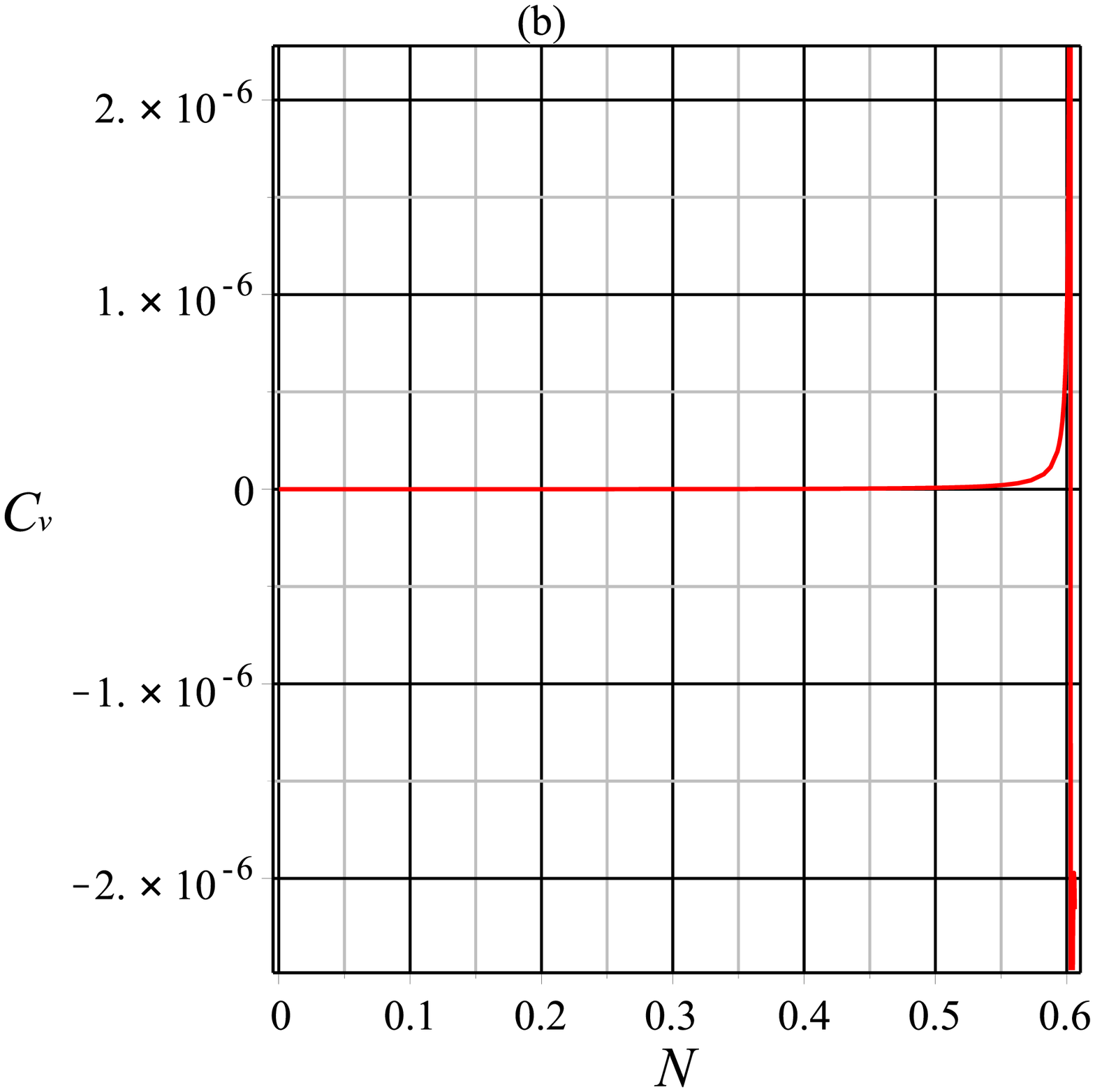}
 \end{array}$
 \end{center}
\caption{Typical behavior of the specific heat in terms of (a) $T$, and (b) $N$ for the unit values of the parameters.}
 \label{figC2}
\end{figure}

Using the above entropy, one can calculate specific heat via,
\begin{equation}\label{eq7.45}
C_{v}=T\left(\frac{d{\cal S}}{dT}\right)_{V}.
\end{equation}
and the numerical analysis  is represented by plots of Fig. \ref{figC2}.\\

\nd
Fig. \ref{figC2} (a) shows the specific heat in terms of temperature. Similar to the previous case, we can see Schottky anomaly appearing here as well. Depending on the number of galaxies there is a minimum temperature for which  specific heat is positive. It means that at low temperatures, the system of galaxies in the brane world model is unstable. We can also see a phase transition which is illustrated by plot of Fig. \ref{figC2} (b). For a big number of galaxies, the system goes to unstable phase. Hence, we conclude that the stable system exists in lower number and higher temperature.\\

\begin{figure}[h!]
 \begin{center}$
 \begin{array}{cccc}
\includegraphics[width=70 mm]{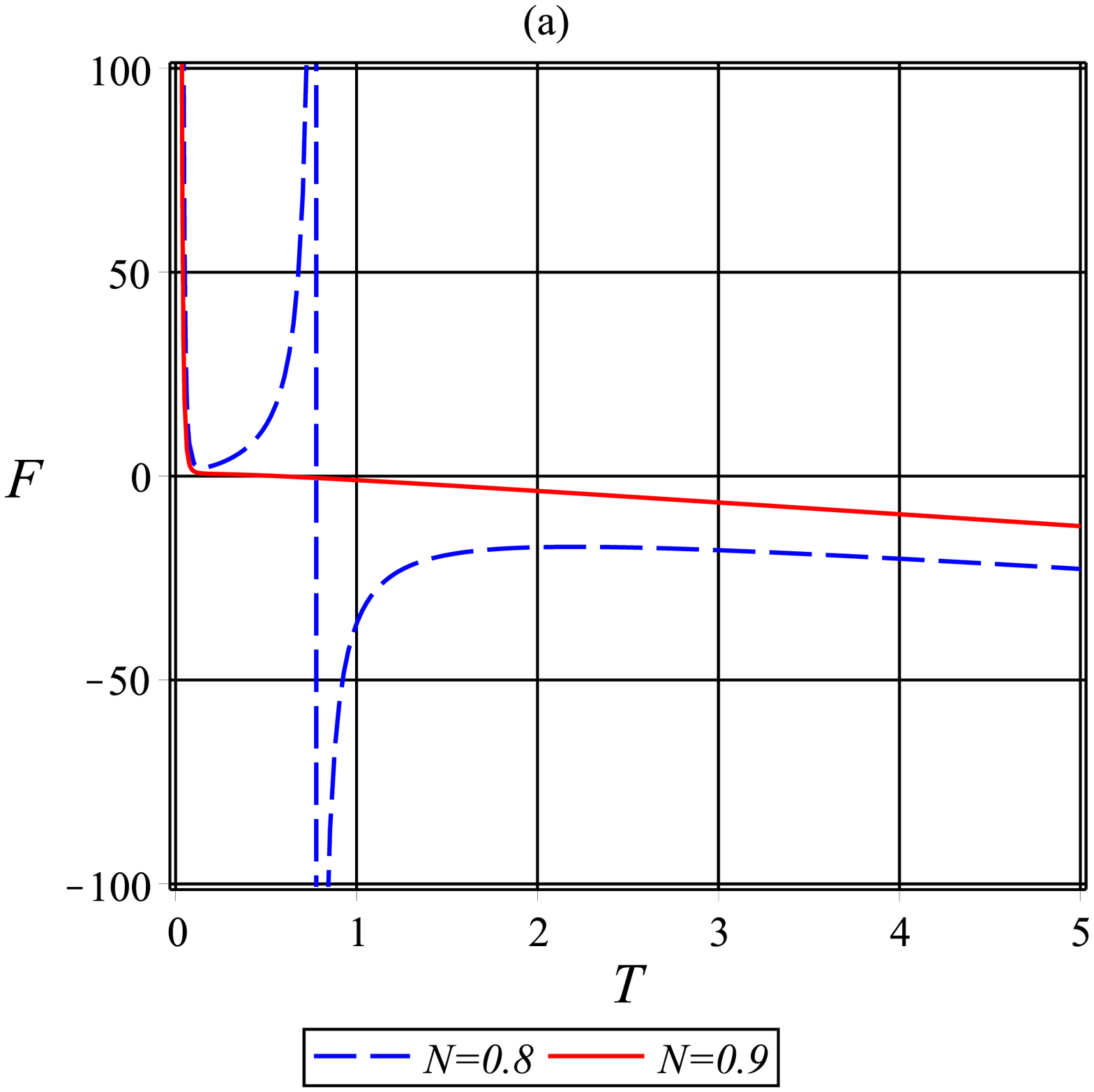}\includegraphics[width=70 mm]{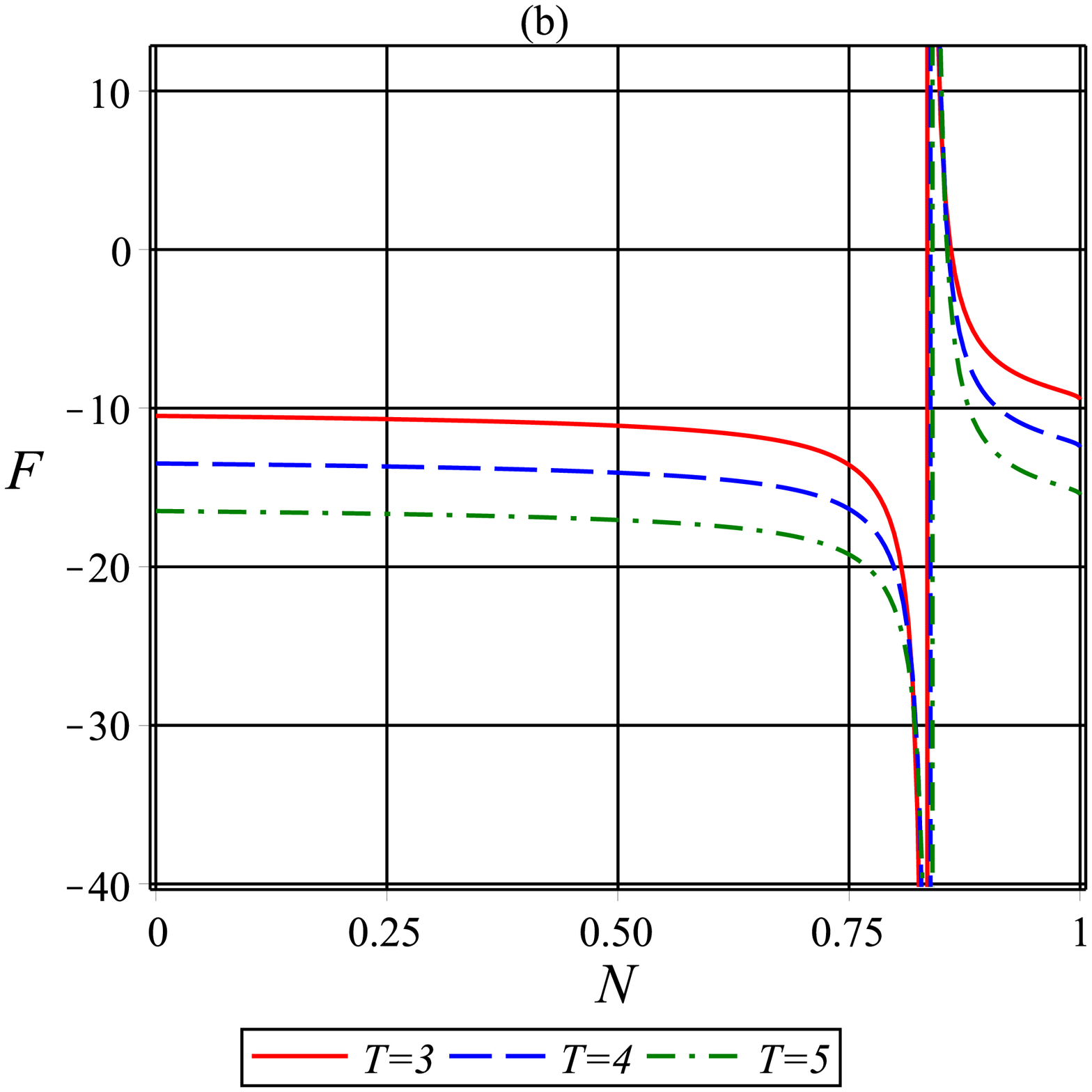}
 \end{array}$
 \end{center}
\caption{Typical behavior of the Helmholtz free energy in terms of (a) $T$, and (b) $N$ for the unit values of the parameters.}
 \label{figF2}
\end{figure}

\nd
Finally, the Helmholtz free energy is calculated as,
\[{\cal F}=-\frac{h(3)\left(C+\frac{(g(3)h(3))^{'}}{g(3)h(3)}\right)+\Phi(3,s)}{\left[ \left[\frac{1}{6}-\frac{28}{9}s\left(1-\frac{24s}{5}+\frac{288s^2}{35}-\frac{32s^3}{7}\right)\right]\left(C+\frac{f^{'}(3)}{f(3)}\right)+\phi_3(s)\right]}\]\[- 3T+\left(\frac{h(3)\left(C+\frac{(g(3)h(3))^{'}}{g(3)h(3)}\right)+\Phi(3,s)}{\left[ \left[\frac{1}{6}-\frac{28}{9}s\left(1-\frac{24s}{5}+\frac{288s^2}{35}-\frac{32s^3}{7}\right)\right]\left(C+\frac{f^{'}(3)}{f(3)}\right)+\phi_3(s)\right]}+3\right)\]
\begin{equation}
\label{eq7.5}
  {\biggl(\frac{1}{18}g(3)\left[ \left[3-56s\left(1-\frac{24s}{5}+\frac{288s^2}{35}-\frac{32s^3}{7}\right)\right]\left(C+\frac{f^{'}(3)}{f(3)}\right)+18\phi_3(s)\right]\biggr)}^{-\frac{1}{3}}
\end{equation}
\nd
In Fig. \ref{figF2} we can see typical behavior of the Helmholtz free energy by variation of $T$ and $N$. Figures \ref{figF2} (a) and (b) show that Helmholtz free energy has asymptotic behavior for the smaller $N$ and $T$. Moreover, larger $T$ and smaller $N$ yield to the negative Helmholtz free energy.

\section{Conclusion and discussion}

In this study, we have considered the brane world model with large distance corrections to study thermodynamic properties of the gravitating systems. We used two different schemes viz;  the Boltzmann and  the Tsallis statistical approaches to calculate the partition function. We found some similarity between the two such as the Schottky anomaly in specific heat analysis. While as in Boltzman statistics, such anomaly appeared in forbidden region, in Tsallis statistics, it happened in the physical system. In both methods we found appropriate domain of the $N$ and $T$ to have physical model. By using the laws of thermodynamics it is found that the brane world model is stable for a lesser number of galaxies. The domain of stability is different for Boltzmann and Tsallis statistics.  Further, it is  found that Boltzmann distribution function is Gaussian.
The results of this paper show
the effect of large distance quantum effects, which opens a new area of study. 
Keeping in view the growing interest in exploring the relevance of quantum effects at large distances, especially due to the rapid developments in the experimental techniques in quantum physics, we propose in our present study the large distance quantum modification to the Newtonian gravity and study the clustering of galaxies.
This is a first of its study of analyzing quantum effects at large distance on the gravitational clustering and structure formation. This paper can thus be treated as a proposal that the quantum effects can reoccur at large distances and can modify Newtonian potential, which in turn modify the partition function along with the corresponding thermodynamics. The results have been plotted and show some unusual effects especially in case of specific heat at bigger number of galaxies and at lower temperatures. This can be a test of stability of the system under certain conditions. There is minimum temperature where the entropy of the system is not well defined and the stability issues with upper limit of temperature is also seen from the plots.   

\nd
This is a first of its study of analyzing quantum effects at large distance on the gravitational clustering. This paper can be treated as a proposal that quantum effects can occur at large distances and thus modify Newtonian potential. Which in turn modify the partition function along with the corresponding thermodynamics. The results have been plotted and show some unusual effects especially in case of specific heat at bigger number of galaxies and at lower temperatures. This can be a test of stability of the system under certain conditions. There is minimum temperature where the entropy of the system is not well defined and the stability issues with upper limit of temperature is also seen from the plots.

\section{Conflict of interest statement}

\nd On behalf of all authors, the corresponding author states that there is no conflict of interest.

\end{document}